\documentclass[preprint,12pt]{elsarticle}

\usepackage[letterpaper]{geometry}

\usepackage{hyperref}
\usepackage{booktabs}
\usepackage{graphicx}
\usepackage[frozencache,cachedir=minted-cache]{minted}
\usepackage[font=small,labelfont=bf]{caption} 
\usepackage{subcaption}

\usepackage{algorithm}
 \usepackage{algpseudocode}
 \usepackage{amssymb}
 \usepackage{lineno}
 \usepackage{tablefootnote}

\renewcommand{\thefootnote}{\alph{footnote}}

\newcommand{\SPEEDUP}{130}

\journal{Computer Physics Communications}

\begin{document}
\emergencystretch 3em

\begin{frontmatter}

\title{Efficient Algorithms for Monte Carlo Particle Transport on AI Accelerator Hardware}

\author[label1]{John Tramm\corref{cor1}}
\ead{jtramm@anl.gov}
\author[label1,label2]{Bryce Allen}
\author[label1]{Kazutomo Yoshii}
\author[label1]{Andrew Siegel}
\author[label3]{Leighton Wilson}
\affiliation[label1]{organization={Argonne National Laboratory},
             addressline={9700 S. Cass Ave.},
             city={Lemont},
             postcode={60439},
             state={IL},
             country={USA}}

\affiliation[label2]{organization={University of Chicago},
             addressline={5801 S. Ellis Ave.},
             city={Chicago},
             postcode={60637},
             state={IL},
             country={USA}}

 \affiliation[label3]{organization={Cerebras Systems Inc.},
             addressline={1237 E. Arques Ave.},
             city={Sunnyvale},
             postcode={94085},
             state={CA},
             country={USA}}
\cortext[cor1]{Corresponding Author}

\begin{abstract}
The recent trend toward deep learning has led to the development of a variety of highly innovative AI accelerator architectures. One such architecture, the Cerebras Wafer-Scale Engine 2 (WSE-2), features 40 GB of on-chip SRAM, making it a potentially attractive platform for latency- or bandwidth-bound HPC simulation workloads. In this study, we examine the feasibility of performing continuous energy Monte Carlo (MC) particle transport on the WSE-2 by porting a key kernel from the MC transport algorithm to Cerebras's CSL programming model. 
New algorithms for minimizing communication costs and for handling load balancing are developed and tested. The WSE-2 is found to run \SPEEDUP~times faster than a highly optimized CUDA version of the kernel run on an NVIDIA A100 GPU---significantly outpacing the expected performance increase given the difference in transistor counts between the architectures.
\end{abstract}

\begin{keyword}
High Performance Computing \sep
    Monte Carlo Particle Transport \sep
    Macroscopic Cross Section Lookup \sep
    AI Accelerators \sep
    Cerebras
\end{keyword}

\end{frontmatter}

\section{Introduction}

As performance gains become more difficult to attain with traditional CPU architectures, hardware accelerators have become increasingly commonplace in high-performance computing (HPC) systems. Much of this diversification has been driven by the need to support artificial intelligence (AI) training, resulting in the widespread adoption of graphics processing units (GPUs) and even more bespoke AI accelerators. Many of these new architectures are highly specialized for deep learning and do not target general HPC workloads. However, some do have innovative characteristics that make them potentially attractive for more traditional HPC. One such recent architecture, the Cerebras Wafer-Scale Engine 2 (WSE-2), is notable not only for its very large physical size but also for its 40 GB of on-chip SRAM available with 1-cycle latency. This characteristic makes it a good fit for HPC workloads that have traditionally been bandwidth- or latency-bound. In the present analysis we study the performance of a historically memory-bound kernel on the WSE-2 AI accelerator.  The kernel, taken from the field of Monte Carlo (MC) particle transport, is an ideal candidate for study given both its critical significance to fission and fusion reactor simulation, as well as its historical difficulty in efficiently utilizing CPU architectures~\cite{Tramm:xsbench}.

To our knowledge, this is the first published study of an MC particle transport algorithm on an AI accelerator. While  several other works have adapted  HPC simulations to Cerebras architectures~\cite{jacquelin2022,vanessendelft2022,sai2023,GB_cerebras}, these have focused on algorithms whose inner loops tend to be composed mostly of dense matrix or regular stencil operations. The present work is based on a highly irregular stochastic algorithm that does not involve matrix operations, is subject to stochastic load imbalances, and requires complex multistage routing of particles through the accelerator's network. Given these added complexities, the present analysis is expected to be highly informative as to the potential for a wider variety of complex and irregular simulation methods to be mapped efficiently onto AI accelerators like the Cerebras WSE-2.

\subsection{Monte Carlo Particle Transport and the Cross Section Lookup Kernel}

Monte Carlo particle transport is a method for simulating the behavior of particles as they move through (and interact with) a medium. The MC process is notable in that it is a direct method that simulates the histories of individual particles, rather than numerically integrating a partial differential equation that models the process. MC is commonly deployed in a variety of scientific and engineering fields because it is both highly accurate and general-purpose, making a minimum of physical assumptions while still being capable of simulating a wide variety of problem types. The downside to MC is that it is both numerically costly (in that many particles must be run in order to reduce uncertainties to acceptable levels) and computationally inefficient (since the MC process is inherently stochastic, resulting in branchy control flow, random memory access patterns, and low natural vector efficiency). Given the method's great utility to several industries (in particular, the simulation of both fission and fusion reactors) and its performance challenges, there has been significant investment in recent years in the development of new MC algorithms for modern  HPC systems~\cite{pragma,hamilton2019ane,Tramm2022,tramm2023,Mckinley2019,Pozulp2023,bergmann2015ane,hamilton2018ane,hamilton2016tans}.

In Monte Carlo neutron particle transport, the macroscopic neutron cross section lookup kernel accounts for the majority of overall program runtime when simulating depleted fuel reactor cores. On CPUs, this single kernel has accounted for up to 85\% of the overall runtime~\cite{Tramm:xsbench}, with similarly high runtime percentages in GPU implementations (for instance, the PRAGMA GPU Monte Carlo code reported that 67\% of runtime was spent performing cross section lookups~\cite{pragma}). Thus, optimization of this kernel is key to optimizing Monte Carlo particle transport in general, and consideration of the macroscopic lookup kernel in isolation is a  valuable exercise that spares us the complexity of having to implement all lower-cost kernels in the MC transport algorithm (e.g., ray tracing, geometry representation, collision physics). In fact, the XSBench mini-app~\cite{Tramm:xsbench} that  represents only the cross section lookup kernel has been commonly used as a stand-in for performance analysis of full-physics Monte Carlo. Thus, we  limit the scope of this paper to consideration of only the macroscopic cross section lookup kernel, leaving implementation of other kernels within the Monte Carlo transport loop as tasks for future research.

The macroscopic cross section (XS) lookup kernel's function within the MC particle transport routine is to assemble statistical distribution data stored in a number of tables, which is then used to generate random samples for a particle's behavior as it moves through a simulated geometry and undergoes different interactions with the materials (e.g., scattering, absorption, escape, fission, etc.). Thus, in order to facilitate stochastic sampling of each event in a particle's history, cross section data must be looked up. This data has several dependencies, including the energy of the neutron, the isotopic composition of the material that it is traveling through, and the temperature of the material. Microscopic cross section data is typically stored in tabular form for each nuclide, with several reaction channels of data stored for each of thousands of different energy levels per nuclide. The energy grids for each nuclide are typically unique, since they are generated  to bound the error when interpolating values outside of the table. Energy spacing within the various nuclide tables also varies, becoming very fine in locations that have sharp resonance features, which are located at different energy levels for different nuclides. Thus, in order to assemble a macroscopic cross section for a particle, a separate lookup operation is required to locate and interpolate microscopic data from each nuclide's grid and then multiply each nuclide's microscopic cross section quantity by that nuclide's density within the material. These nuclide-wise quantities are then summed together to form the final macroscopic cross section value that can be used when sampling the particle's stochastic behavior, as

\begin{equation}
    \label{eq:macro_xs}
    \Sigma_r(e) = \sum_{n}^{material} \sigma_{r,n}(e)  \rho_{n},
\end{equation}
\noindent
where $\Sigma_t$ is the macroscopic cross section at energy $e$ for reaction channel $r$, $n$ is the nuclide within a material, $\sigma_{n}$ is the microscopic cross section, and $\rho_{n}$ is the density of that nuclide within the material, and the first character in the right-hand side of \autoref{eq:macro_xs} represents a summation over nuclides. Multiple cross section reaction channels are stored at each energy level (e.g., total, absorption, scattering, fission, nu-fission), and typically all are computed for each cross section lookup operation. Note that including temperature dependence does not fundamentally alter the inner loop structure, so we exclude it for simplicity.

A simplified pseudocode is given in Algorithm~\ref{alg:kernel}, which corresponds to a single lookup for a single particle at energy $e_{particle}$. In both XSBench and the present implementation, we add an outer loop over many independent particles with randomized energy levels, each of which requires an energy lookup, as is typical of the way this kernel is expressed as part of the ``event-based'' formulation of the Monte Carlo particle transport loop~\cite{brown1984pne}.

\begin{algorithm}
  \caption{Simplified macroscopic cross section lookup kernel
    \label{alg:kernel}}
  \begin{algorithmic}[1]
   \State $\vec\Sigma \gets 0$ \Comment{Macroscopic XS vector}
        \For{nuclide $n$ in material}
            \State lower bound $\gets$ binary search for $e_{particle}$ in $\vec e_n$
            \State $e_{n,low} \gets \vec e_{n}[$lower bound$]$
            \State $e_{n,high} \gets \vec e_{n}[$lower bound$ +1]$
            \State $f \gets \frac{e_{particle} - e_{n,low}}{e_{n,high} - e_{n,low}}$ \Comment{Interpolation factor}
            \For{channel $r$ in $\vec\Sigma$ }
                \State $\sigma_{n,r, low} \gets \vec \sigma_{n}[$lower bound$][r]$
                \State $\sigma_{n,r, high} \gets \vec \sigma_{n}[$lower bound$ +1][r]$
                \State $\sigma_{n,r,e_{particle}} \gets \sigma_{n,r, high}- f (\sigma_{n,r, high} - \sigma_{n,r, low})$
                \State $\vec\Sigma[r] \gets \vec\Sigma[r] + \rho_n \sigma_{n,r,e_{particle}}$
            \EndFor
        \EndFor
  \end{algorithmic}
\end{algorithm}

The XSBench mini-app is a simple representation of Algorithm \ref{alg:kernel} for depleted fuel nuclear reactor simulations that typically track hundreds of nuclides in the fuel material. While fresh fuel typically contains only a handful of nuclides (oxygen-16, uranium-235, uranium-238), hundreds of fission products and actinides (as well as their subsequent decay chains) are produced over time that must be tracked in a simulation. While Monte Carlo particle transport methods can be used to simulate a wide variety of problems beyond fission reactors (e.g., fusion reactor design, medical dosimetry, shielding), for the purposes of grounding our analysis, the present analysis will consider problem configurations typical of full-core nuclear reactor simulations featuring depleted fuel.

\subsection{Cerebras WSE-2 Hardware Architecture Overview}

The Cerebras WSE-2 architecture differs from traditional HPC architectures in a number of ways. The first, and most notable, is the scale of the chip. While a modern GPU such as the NVIDIA A100 has a die area of 826 mm$^2$, the Cererbas WSE-2 is a ``wafer-scale'' chip that is scaled to utilize an entire silicon wafer at once, thus having a die area of 46,225 mm$^2$. In fact, one might argue that term ``chip'' does not accurately describe the WSE-2, since the etymology of this term stems from the process of breaking off a small chip from a large silicon wafer. In addition to its greatly increased scale, the WSE-2 differs greatly from CPU or GPU architectures in that it is customized for deep learning, featuring specialized instructions and hardware to handle the 16-bit matrix operations that are common to these workloads. While GPUs also feature specialized AI instructions, they also  provision significant die space for resources devoted to graphics processing and more general-purpose workloads. The WSE-2 also does not not support 64-bit floating-point arithmetic;  32-bit floating-point (single-precision FP32) is the highest precision supported.

\begin{figure}[h!]
    \centering
    \includegraphics[width=0.5\columnwidth]{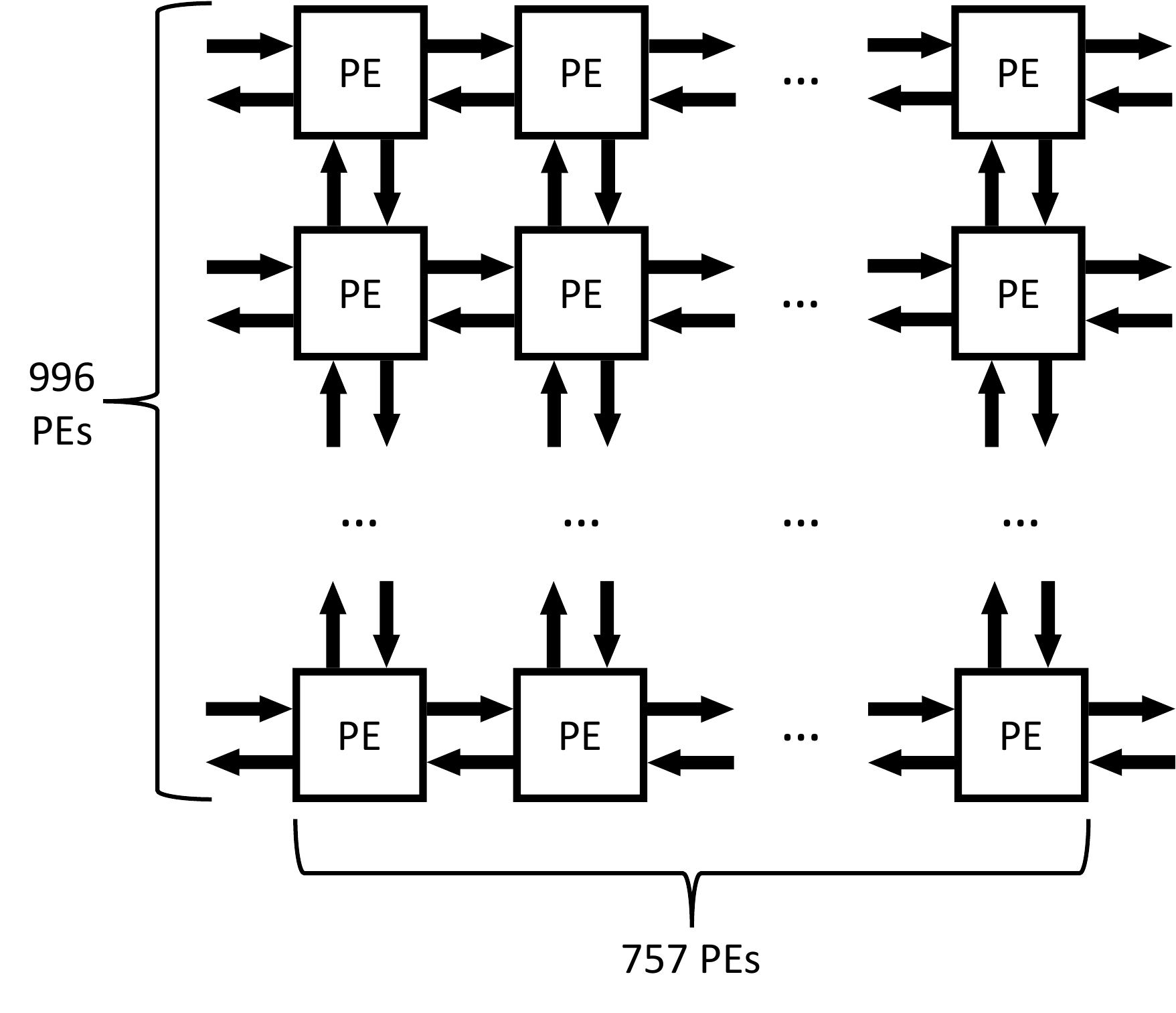}
    \caption{Diagram of WSE-2 architecture.}
    \label{fig:WSE-2}
\end{figure}

Another unique feature of the WSE-2 is its lack of off-chip memory. The WSE-2 architecture does not utilize any dynamic random access memory (DRAM). Rather, all memory comes in the form of single-cycle latency static random access memory (SRAM), of which about 40 GB is available on the WSE-2. This singular feature makes the architecture a potentially good fit for HPC simulation kernels that are memory bandwidth or latency bound, since the theoretical bandwidth of the WSE-2 is staggering at 20 petabytes/second (compared with the A100 GPU, which has 1.5--2.0 terabytes/second of bandwidth, depending on the specific model of A100). However, the low latency and high bandwidth come with a caveat---namely, that the 40 GB of SRAM is distributed across 850,000 cores. Furthermore, the WSE-2 is not a shared-memory architecture. Each compute core (called a ``processing element'' (PE) on the WSE-2) has only 48 kB of local SRAM, with no ability to abstractly access any other memory spaces. Communication of data between PEs must be done manually by the application developer via a distributed-memory message-passing model. A simplified diagram of the WSE-2 architecture, shown in \autoref{fig:WSE-2}, shows how the WSE-2's PEs are arranged in a 2D grid, with each neighbor interconnected with one another.

Another important characteristic of the architecture shown in \autoref{fig:WSE-2} is that the latency between PEs can be as low as 1 cycle. Additionally, communication between PEs is typically done asynchronously, allowing for natural pipelining of data between PEs, where the already low communication latency can theoretically be masked. Thus, while the architecture can be programmed as a large 2D network of cores (much in the way HPC architectures have historically approached distributed-memory parallelism via MPI, though with major caveats), it can also be treated as a dataflow architecture, where each PE is configured to perform only a single small task on an object before sending it to the next PE. We also note that unlike typical network architectures such as Ethernet or InfiniBand that are programmed with MPI, the WSE-2 network is strictly a neighbor-to-neighbor network. Messages cannot be passed abstractly between any processor in the grid to any other processor. Rather, they must be routed manually by the programmer between neighbors. In the Cerebras Software Language (CSL) used to program the WSE-2, a few primitive operations are provided abstractly, such as reduction operations across rows or columns of the WSE-2, but in general most communication schemes must be programmed manually. Given the limited memory capacity of each PE (only 48 kB), sophisticated communication runtimes (e.g., as in MPI) are not practical. Rather, message passing is done in a very low-level manner, where the PE's router and the various limited hardware message queue resources on each PE must be explicitly configured and managed. 

Thus, while the architecture is highly attractive due to the 40 GB of 1-cycle latency SRAM, decomposing data into 48 kB subdomains (and coordinating communication between subdomains using a low-level message-passing model) may present significant challenges. The adaptation of any computational HPC kernel onto the WSE-2 will therefore require new communication algorithms and optimization techniques to facilitate decomposition into kilobyte-scale subdomains.

\subsection{Cross Section Lookup Kernel on the WSE-2}

Our goal is to reproduce the cross section lookup kernel as defined in XSBench for the WSE-2 using the Cerebras SDK and the CSL programming model\footnote{Our final CSL implementation (and all other code used to support this paper) is open source and publicly available at~\cite{mc_csl_zenodo}.}, with a few simplifications. XSBench represents a realistic lookup pattern that involves lookups from materials with few nuclides (such as the coolant) as well as fuel materials (with hundreds of nuclides) using realistic distributions of lookup frequencies taken from real Monte Carlo simulations. However, recent work in the field of GPU Monte Carlo has often found that it is optimal to partition lookups into two separate event-based kernels~\cite{Tramm2022,pragma,hamilton2019ane}. The first kernel handles only the expensive fuel lookups, while the second kernel handles the comparatively much cheaper lookups for all other materials in the simulation.  
Thus, for the sake of simplicity, our implementation only considers lookups in a single depleted fuel material.

Similar to XSBench, we also utilize randomly generated synthetic cross section data, which is justified  given that we care only about mimicking the computational patterns and are not concerned that the resulting macroscopic cross section data has no physical meaning. While the number of nuclides (250) in our implementation is realistic for depleted fuel, real cross section data has a variable number of energy gridpoints per nuclide, ranging from roughly a thousand to over 100,000. In XSBench and our implementation for the WSE-2, we simplify this to a constant 10,000 gridpoints per nuclide, which corresponds with the approximate average number in a real depleted fuel problem~\cite{Tramm:xsbench}. We also utilize 32-bit data for both the energy grid and  underlying cross section data, as is done in the 32-bit version of XSbench~\cite{tramm2023}. 

\begin{table}[h]
\centering
\caption{Simplified cross section lookup parameters representing a depleted fuel material in a nuclear reactor simulation.}
\label{tab:xs_params}
\begin{tabular}{@{}ll@{}}
\toprule
MC Cross Section Kernel Parameter & Value \\ \midrule
Nuclides & 250 \\
Energy gridpoints per nuclide & 10,000 \\
Cross section reaction channels & 5 \\
Bytes per 32-bit value & 4 \\ \midrule
Total cross section + energy data & 60 MB \\ \bottomrule
\end{tabular}
\end{table}

Given the parameters listed in \autoref{tab:xs_params}, it is clear that cross section data cannot be replicated on each PE (which has only  48 kB of local memory); it must be decomposed across many PEs of the WSE-2.

\section{Compute Kernel Optimization}
\label{sec:single_pe_opt}

We begin by optimizing the basic cross section lookup computation, assuming temporarily that all required data fits within a single processing element. The naive implementation using the Cerebras CSL programming model is  straightforward and requires roughly the same number of lines of code as the standard C implementation. While  additional boilerplate code in CSL is required for defining where within the WSE-2's grid the kernel will launch and for moving data between the host and device, in general the code  complexity is similar to that of  most other device-oriented programming models (e.g., CUDA, SYCL, HIP, OpenMP). Listing \ref{listing:xs_kernel} gives an example of the basic kernel definition in CSL.

\begin{listing}[h!]
\begin{minted}[%
 breaklines,
 mathescape,
 linenos,
 numbersep=5pt,
 frame=single,
 xleftmargin=10pt,
 ]{zig}
fn calculate_xs() void {
  for (@range(i16, n_particles)) |p| {
    var e: f32 = particle_e[p];
    for (@range(i16, n_nuclides)) |n| {
      // Perform binary search
      var lower: i16 = bsearch(n, e);

      // Compute Interpolation factor
      var e_lower  : f32 = nuclide_energy[n, lower];
      var e_higher : f32 = nuclide_energy[n, lower + 1];
      var f : f32 = (e_higher - e) / (e_higher - e_lower);

      // Interpolate and store XS to particle
      for (@range(i16, n_xs)) |xs| {
        var xs_lower  : f32 = nuclide_xs[n, lower, xs];
        var xs_higher : f32 = nuclide_xs[n, lower+1, xs];
        particle_xs[p, xs] += densities[n] * (xs_higher - f * (xs_higher - xs_lower) );
      }
    }
  }
}
\end{minted}
\caption{Simplified Monte Carlo cross section lookup kernel implemented in CSL. For brevity, multi-dimensional array indexing is shown, although our final kernel implementation uses 1D indexing.}
\label{listing:xs_kernel}
\end{listing}

However, a number of potential optimizations are possible at this scale. Several of these focus on the linear interpolation operation. This operation is needed because the cross section data is stored exactly at each energy gridpoint location. In order to determine the correct cross section values between energy gridpoints, a simple linear interpolation operation is performed, as in \autoref{eq:linear_interp}, where $f$ is the computed interpolation factor, $e_{low}$ and $e_{high}$ are the lower and higher bounding energy grid levels, respectively, and $e$ is the particle's current energy level. With $f$ computed, it can then be used to interpolate each of the microscopic cross section reaction channels.

\begin{equation}
    \label{eq:linear_interp}
    f = \frac{e_{high} - e}{e_{high} - e_{low}}
\end{equation}

This is an expensive operation on the WSE-2 because  32-bit floating-point division operations take around 50 cycles to complete. We consider several optimizations. The first potential optimization is to simply replace the 32-bit division operation with a 16-bit division operation. While this will result in some small loss of accuracy of the interpolation factor, given the uncertainties in the measured cross section data, the loss of precision in this operation is not expected to impact the accuracy of a real simulation. One complication of this optimization is that the operands and result will need to be converted between FP16 and FP32 formats, which is not free, but may still be cheaper than the savings from the reduced-precision division operation.

The second potential optimization uses a stochastic treatment for the interpolation operation. This treatment functions by generating a sample at a random energy level, $s$, from a uniform distribution between the two bounding energy gridpoints ($e_{low}$ and $e_{high}$), and then comparing this sample with the particle energy $e$ in order to determine which of the bounding datasets to select. If the particle's energy is above the sample, select the higher gridpoint's cross section data to use. If the particle's energy is below the sample, select the lower gridpoint's data. This operation is  statistically identical to linear interpolation, although it does have the downside of adding in a very small amount of additional variance into the overall simulation.\footnote{Because of the use of randomized (synthetic) cross section data in our CSL implementation, we cannot  accurately gauge the impacts of this change on variance in a realistic simulation using real data. To ensure that the use of stochastic interpolation is a valid optimization, we therefore implemented the stochastic interpolation scheme into the full-physics OpenMC Monte Carlo particle transport code~\cite{romano2015ane1} and tested each method on a simulation benchmark problem of a realistic depleted full-core small modular reactor. To amplify the impacts of this change, we disabled use of $S(\alpha,\beta)$ calculations in the thermal neutron regions as well as probability tables in the unresolved resonance range. For a simulation with 12.5 million active batch particles in total, we found that OpenMC produced a k-eff eigenvalue of $1.00498 \pm 0.00026$ using normal linear interpolation and $1.00504 \pm 0.00027$ using stochastic interpolation. Thus, both solutions were well within statistical uncertainty, and the difference in the magnitudes of the uncertainties themselves was negligible, meaning that the stochastic interpolation strategy can be considered a valid optimization in our CSL implementation.} We note that on CPU and GPU architectures, this is not a useful optimization, because both data points are typically located on the same (or adjacent) cache lines, such that accessing both sets is not likely to result in a cache miss. Furthermore, usage of a pseudorandom number generator (PRNG) introduces additional overhead, which depending on the PRNG algorithm may involve a floating-point division operation. However, the WSE-2 hardware actually has specialized PRNG hardware for producing random variates efficiently, making stochastic interpolation potentially much more attractive.

A final optimization leverages the vector units on the WSE-2 PE hardware to perform the inner loop over reaction channels in Algorithm \ref{alg:kernel}. This is accomplished by utilizing intrinsic vector functions in CSL, which the compiler does not currently utilize if operations over vectors are expressed more plainly in the form of typical iterative \texttt{for} loops. We implement this optimization only for the case where stochastic interpolation is used, since this can be done easily using only a single fuse multiply-add (FMA) vector instruction (\texttt{@fmacs}, in CSL).

All three of the proposed optimizations are implemented into a single-PE implementation in CSL for testing on a single PE of a WSE-2 machine. Since a realistic problem size (using hundreds of nuclides and thousands of gridpoints per nuclide) cannot fit onto a single PE, we select a problem size that corresponds to the subdomain a single PE might possess if run in a domain-decomposed manner. Our test problem features a single nuclide, with 161 energy gridpoints, 5 cross section reaction channels, and with 100 particles. Thus, this portion of our analysis does not account for communication costs, as will be discussed later in the paper.

We collect results by running on a Cerebras CS-2 Wafer-Scale Cluster that features several CPU nodes, along with two CS-2s, each with a single WSE-2 chip. All runtime measurements in this paper are made by running on a single WSE-2 of a CS-2. Using the Cerebras SDK, each WSE-2 exposes up to $750 \times 994$ user programmable PEs (i.e., 745,500 PEs). While the WSE-2 hardware actually contains about 850,000 PEs in total, some additional rows and columns around the user-space PEs are reserved for memory movement operations (to facilitate abstractions for moving data to/from the host) and other system functions. CSL allows a programmer to define smaller grids than the maximum allowed by the hardware, or even for running on just a single PE at a time. To determine the runtime of the kernel, the CSL language exposes hardware clock cycle timer data that can be queried and saved over the runtime of a kernel and reported back to the host. Thus, the total wall time of a kernel can be computed by determining the maximum number of cycles used by any PE during the kernel and dividing it by the clock rate of the WSE-2 (850 MHz). While some degree of thermal throttling will occur, the WSE-2 implements throttling by injecting ``nop'' commands rather than by adjusting the clock speed itself, such that any thermal ``nop'' cycles are included when measuring kernel cycle counts. This method of measuring kernel runtime performance by recording clock cycles is used throughout this paper whenever runtime data is reported and is the standard method for recording performance data on Cerebras machines~\cite{jacquelin2022}.

Performance results for the single-PE optimization strategies, shown in \autoref{tab:pe_optimize}, indicate that the switch to stochastic interpolation using WSE-2's PRNG hardware has a significant performance impact, resulting in about a 65\% overall kernel speedup. 
The FP16 division option also provides modest speedup (about 14\%), but given that the stochastic interpolation optimization was much more significant, we choose the latter approach for the final kernel implementation. Use of vector intrinsic operations for the inner loop over reaction channels also netted a small benefit, reducing cycle counts from 281 down to 250, so this optimization (in addition to stochastic interpolation) was selected for our final implementation. With these optimizations in place, the majority of kernel cycles are spent performing binary search operations, such that further optimizations to the kernel would likely need to target lower-level optimizations of the binary search routine.

\begin{table}[h!]
\centering
\caption{Cross section lookup kernel optimizations for single WSE-2 PE. Cycle counts are per-particle.}
\label{tab:pe_optimize}
\begin{tabular}{@{}lll@{}}
\toprule
 & \begin{tabular}[c]{@{}l@{}}Cycle\\ Count\end{tabular} & \begin{tabular}[c]{@{}l@{}}Speedup\\ Over\\ Baseline\end{tabular} \\ \midrule
Baseline & 463 & - \\
FP16 & 405 & 14\% \\
Stochastic Interpolation (Software) & 399 & 16\% \\
Stochastic Interpolation (Hardware) & 281 & 65\% \\
Stochastic Interpolation (Hardware) + Vectorization & 250 & 85\% \\ \bottomrule
\end{tabular}%
\end{table}

\section{Cross Section Data Decomposition}
\label{sec:decomp}

With the single-PE kernel implementation well optimized, we now extend our analysis to use a realistically sized cross section dataset decomposed across many PEs of the WSE-2's grid. Since the target simulation requires at least 60 MB of data (as shown in \autoref{tab:xs_params}) and since each PE has only 48 kB of local memory, decomposition to at least $\mathcal{O}(1000)$ PEs is required. Cross section data in our simplified kernel has three dimensions (nuclide, energy, reaction channel) that are available for decomposition.  This leads to two high-level strategies for decomposing data. In the first strategy, data is decomposed in a static manner, with particles flowing through the PEs as needed. The second strategy leaves the particles in place and flows the cross section data through the network. In general, fixing cross section data and moving particles is optimal when particle object sizes are  small, and/or when using a small number of particles per PE. Conversely, it is more efficient to fix the particles in place and flow the cross section data through when using many particles and/or when particle object sizes  are very large, such that the total size of particle objects distributed across the network becomes less than the total size of cross section data. Given the simplicity of our present analysis and the use of small particle objects (being  composed only of an energy field and five cross section reaction fields), we focus on the case where cross section data is statically decomposed and particles are moved through the network. Analysis of alternative cross section data movement communication patterns is left for future work.

The concept of decomposition of cross section data has also been considered for improving locality in CPU-based MC simulations (e.g., via ``energy banding''~\cite{siegel2014}), which functioned by limiting particles to certain energy ranges of data local to a processor. When particles fall below the cutoff for that energy range, they are either transmitted to another processor that stores that data, or the particles are buffered until the next energy range of cross section data is loaded. The decomposition of data across PEs of a WSE-2 is similar, although it requires an even finer-grained approach given the extremely limited memory resources (48 kB) of each WSE-2 PE. 

To this end, we have developed a scheme wherein each row in the PE grid is dedicated to an energy band, while each column is dedicated to a single nuclide. We do not decompose reaction channels because access to these fields is often contiguous and the binary search cost of locating the correct energy grid point is thus amortized, so it is unlikely to provide a net performance benefit. 

This approach is expected to be reasonably efficient when employed in a production Monte Carlo particle transport application, provided that an event-based mode is used. Under these conditions, a kernel invocation would represent a single cross section lookup event for all particles that are located within a fuel material within a reactor geometry.

\section{Communication Patterns}

\subsection{Starting Conditions}

Given a data decomposition scheme, we  now consider how particles (tasks) will traverse the WSE-2's 2D network so that they reach the needed data given a particle's current energy level.
Our implementation initializes all PE's with the same number of starting particles, though with the energy distribution of particles fully randomized. A key implication of this starting condition is that after particles are sorted into bins by energy within a column (as will be discussed in \autoref{sec:sorting}), the per-PE load distribution of particles during the cross section accumulation phase (as will be discussed in \autoref{sec:row_nuclide_accum}) becomes imbalanced.
This starting particle distribution is reasonable if the kernel is invoked within a well-developed event-based simulation, where particles have already been sampled and undergone several events after which they are randomly distributed in energy space.\footnote{This assumption would be violated during the first kernel invocation of each Monte Carlo batch in a real eigenvalue transport simulation because the fission particle distribution follows the Watt spectrum and is therefore clustered in the high-energy bands. However, as has been shown with many GPU event-based implementations~\cite{Tramm2022}, typically several thousand events will be undergone by the particle within its lifetime in a reactor model; and when fresh particles are mixed in after some die early, we would expect that after the first few kernel calls the thousands of subsequent kernel calls will experience a generally random particle energy distribution. Thus, we believe that optimization of the kernel given these starting particle parameters would provide an accurate performance picture for the kernel if it were implemented within a full Monte Carlo particle transport loop.}

\subsection{Energy Sorting}
\label{sec:sorting}

With a random distribution of particle energies as the starting condition for any given column in the PE grid, the first step is to sort all the particles in the column into the correct energy band (row). To accomplish this, we implement a routine wherein each column of PEs will form a 1D communication pattern, with each stage of communication involving PEs exchanging particle data with their upper and lower neighbors. For the highest and lowest PEs in the column, particles will not be transmitted above or below them, respectively, since these rows represent the highest and lowest energy bands possible. While we assume each PE will start with the same number of particles, the randomized energy distribution means that message lengths between PEs in the column must be variably sized, because after several hops, particle buffers will naturally diminish in length as PEs claim particles from buffers into their own energy band. 

\begin{figure}[h!]
    \centering
    \includegraphics[width=0.6\columnwidth]{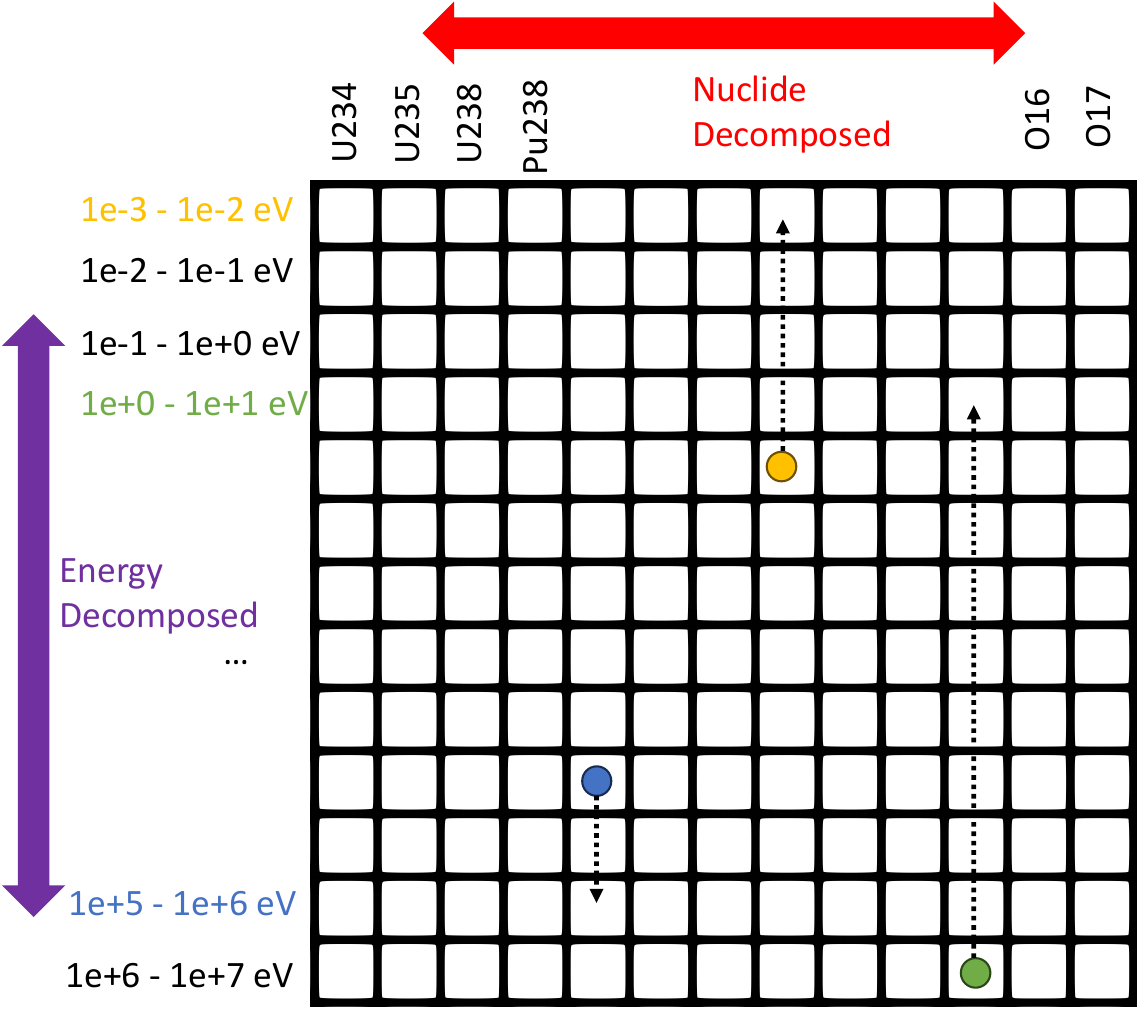}
    \caption{Diagram showing the energy column sorting process. Each square represents a single PE in the WSE-2 grid, with each dot representing a particle. Particles are color coded according to their energy band. While only three particles are shown, in reality each PE may be processing many particles at once. Particles move to their destination rows by moving one row at a time, being evaluated to determine whether they are in the correct energy band and then  being either claimed or transmitted again.}
    \label{fig:column_sort_diagram}
\end{figure}

To determine when communication is complete, we break this pattern into communication iterations, wherein particles are sent and received between nearest neighbors once per iteration. While the exact number of iterations needed to move all particles to their energy band rows is unknown, we can bound this value by considering the worst case where a particle starting at the bottom row needs to migrate to the top row, which for a column of height $h$ will require $h-1$ communication iterations. This inevitably results in unnecessary communication overhead. A more optimized strategy might utilize the CSL ``tally kernel module,'' which is a library-based method for coordinating termination criteria between many PEs. For the present case, however, we use the simplified communication model that  performs $h-1$ communication iterations and otherwise does not involve any coordination between PEs to determine whether termination criteria are met. A simplified diagram depicting this process is shown in \autoref{fig:column_sort_diagram}.

\begin{figure}[h!]
    \centering
    \includegraphics[width=0.7\columnwidth]{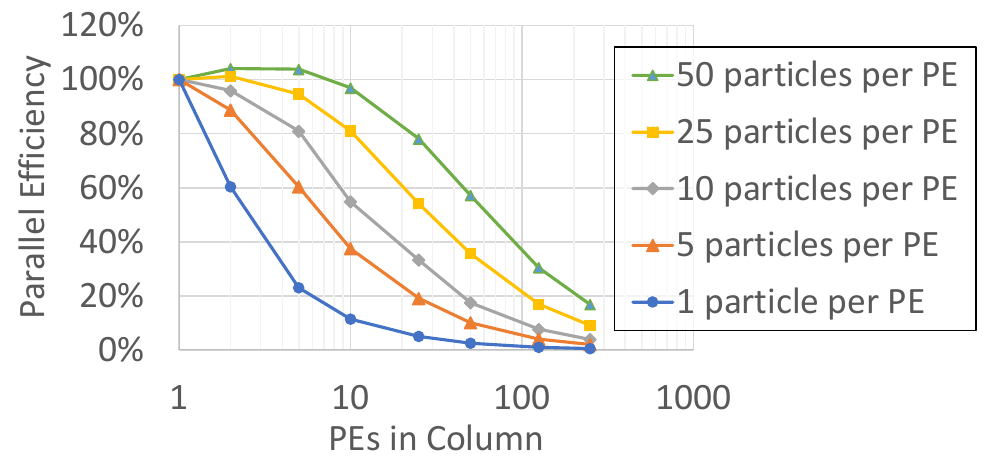}
    \caption{Weak scaling energy decomposition study across a column of PEs. The problem size involves one nuclide with 1,000 energy gridpoints per energy level (row) and five cross section reaction channels. Problem size per PE is fixed with $n$ particles, with several studies run for various quantities of $n$.}
    \label{fig:col_weak}
\end{figure}

We begin our column-sorting performance analysis with a weak scaling study along a single column of the PE grid. 
Our baseline weak scaling problem uses 1 nuclide for the column, using 1,000 energy gridpoints and 5 cross section reaction channels. Since our communication pattern requires each PE to execute a task (e.g., processing of at least a single particle for a single nuclide), we pin the problem size per PE as the number of starting particles per PE, $n$. The sensitivity in performance to $n$ is also investigated. Weak scaling efficiency is then calculated separately for each study of $n$ by comparing the number of cycles per PE per particle against the case where only 1 PE is used for that value of $n$. Results for this single-column study, shown in \autoref{fig:col_weak}, show that the communication costs are significant for more than 10 PEs in a column for all cases, and for as few as two PEs if the particle count per PE is low. While these results may at first appear to indicate impractically high communication costs,  note that the sorting costs in this study are amortized over the lookup of only a single nuclide. Adding more nuclides to the problem should amortize communication costs such that the relative cost of sorting may become small compared with the cost of doing useful cross section lookup work (as will be studied in \autoref{sec:full_comms} when 2D decomposition results are presented).

We also performed a strong scaling study for the column-sorting algorithm. In strong scaling, the global problem size is fixed; and as we add PEs to the row, we spread out the static load (particles)  among more processors. This type of scaling is fairly unnatural on the WSE-2, given that it is difficult to create a meaningful problem with enough work for hundreds of PEs that can also be fit within only the memory resources of a single PE. We fix the number of particles at 100, the number of nuclides at 1, the energy gridpoints per nuclide at 800, and the  cross section reaction channels at 5. Note that as we add rows to the column, the number of gridpoints stored for that nuclide per PE will also decrease due to energy decomposition, which has the positive effect of slightly reducing the number of binary search operations that must be performed as more PEs are used.

\begin{figure}[h!]
    \centering
    \includegraphics[width=0.7\columnwidth]{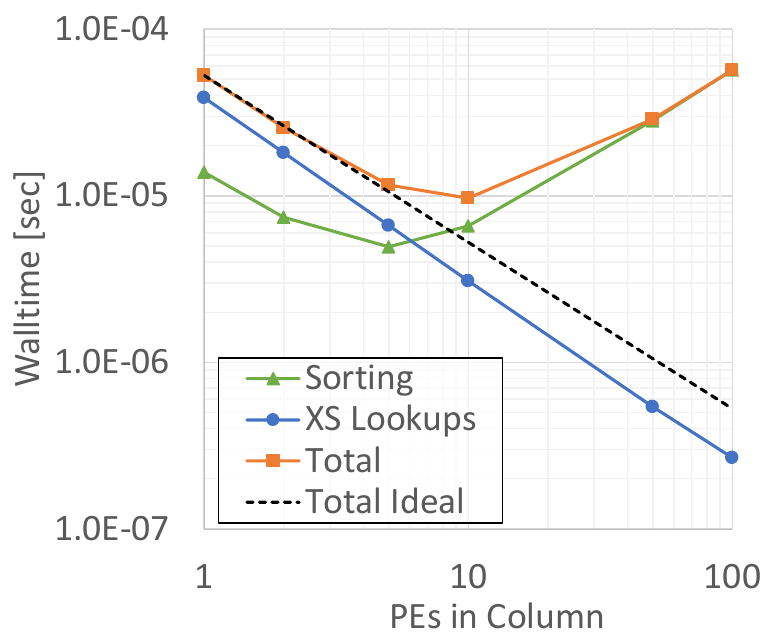}
    \caption{Strong scaling energy decomposition study across a column of PEs. The global problem size is fixed with 100 particles and 1 nuclide, with 800 energy grid points and 5 cross section reaction channels per nuclide.}
    \label{fig:col_strong}
\end{figure}

The results of our strong scaling analysis, given in \autoref{fig:col_strong}, show that for greater than 10 PEs in the column, the sorting communication costs become too expensive for further performance gains. Notably, the sorting costs are non-zero when only a single PE is present, since the PE must analyze all 100 starting particles to determine whether they are in the local energy band and copy them into a separate buffer, which has a significant startup cost associated with it. As PEs are added to the column, the cost of this initial sorting operation is diminished due to each PE starting with fewer particles---resulting in an overall efficiency gain that outweighs the added costs of transmitting particles between PEs. Once approximately 10 PEs are used, however, the communication costs tend to surpass the time it takes to do the actual lookups, after which point use of more PEs results in an increase in runtime. This strong scaling analysis is useful in that it makes clear the benefit to decomposing to at least 10 energy bands, even in the case of a very small problem size. It also makes clear that decomposition past 10 energy bands is not efficient, although we will investigate empirically in \autoref{sec:full_comms} whether the costs of sorting end up being small enough that overall performance is not impacted.

\subsection{Nuclide Accumulation}
\label{sec:row_nuclide_accum}

\begin{figure}[h!]
    \centering
    \includegraphics[width=0.6\columnwidth]{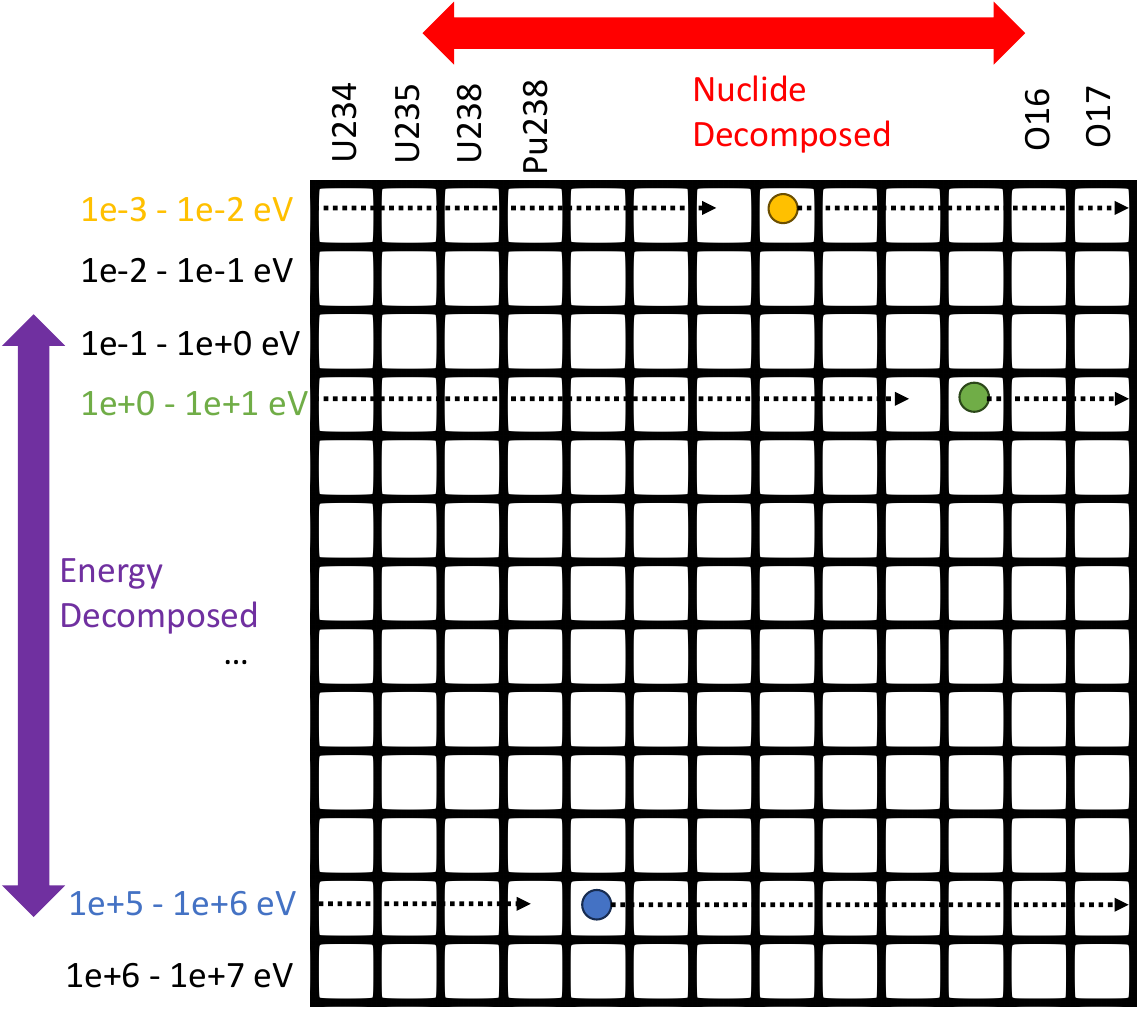}
    \caption{Diagram showing the ``round-robin'' row exchange process for accumulating nuclide data into each particle. Each square represents a single PE in the WSE-2 grid, with each dot representing a particle. While only three particles are shown, in reality each PE may be processing many particles at once. Particles visit each PE in the row once, accumulating nuclide cross section information for one or more nuclides at each visit.}
    \label{fig:row_round_robin_diagram}
\end{figure}

With particles sorted into their appropriate bands, we now must solve the problem of ensuring that each particle is able to access data for every nuclide within the material. As described in \autoref{sec:decomp}, our strategy is to decompose nuclides over columns. Once a particle has arrived in the correct energy band (row) after sorting, the particle must accumulate a contribution to its macroscopic cross section value from each nuclide by traveling to each PE in its row. During its visit to each PE, the particle will perform a lookup and accumulation kernel for all nuclides that the PE owns. We implement this communication pattern in terms of a ``round-robin'' 1D neighbor exchange, where at each communication step a PE will send outgoing particles to its right neighbor and receive incoming particles from its left. Our implementation allows for variable-length messages to be passed between neighbors, with buffer lengths communicated on the fly before actual particle data is transmitted at each communication phase. Periodic boundary conditions are manually implemented such that particles being transmitted from the rightmost boundary PE will be received by the leftmost boundary PE to continue on its round-robin traversal. For a row of width $w$, each particle will therefore make $w-1$ hops until it has visited every PE in the row. A simplified diagram of this process is shown in \autoref{fig:row_round_robin_diagram}.

\begin{figure}[h!]
    \centering
    \includegraphics[width=0.7\columnwidth]{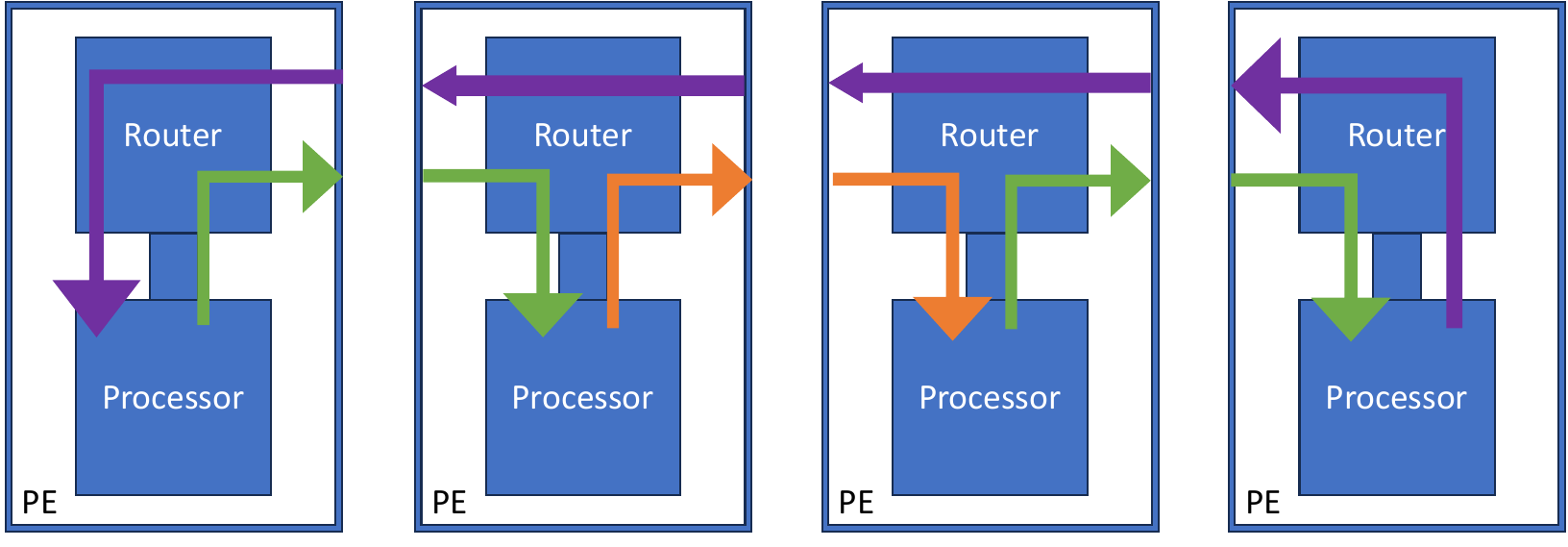}
    \caption{Diagram showing how the 1D ``round-robin'' row exchange process (with a periodic boundary condition) is implemented. Routers are preconfigured to apply specific directional information to messages with specific color tags. Particles visit each PE in the row once, accumulating nuclide cross section information for one or more nuclides at each visit.}
    \label{fig:row_routers}
\end{figure}

We will not provide a detailed description of the implementation of the communication patter in CSL, since this would involve a full explanation of the intricacies of the CSL programming model. However, we provide a cursory overview of some of the hardware characteristics and the basic concepts of message passing on the WSE-2 using CSL. \autoref{fig:row_routers} shows our the basic strategy, where each PE in the row had its router configured to pass messages of different colors in order to accomplish the one-dimensional and one-direction message-passing scheme. Each PE is composed of two disjoint elements: the router and the processor. The router operates independently from the processor, and the processor cannot view any messages unless they are first sent down the ``ramp'' interconnecting the two elements. As the diagram indicates, interior PEs have ``even'' and ``odd'' configurations such that adjacent PEs match colors. For example, the second PE passes to its right on color orange and receives from the left on  color green, while the third PE does the opposite (sending on green and receiving on orange). This diagram also shows how periodic boundary conditions are implemented, with interior PEs simply forwarding the periodic color (purple) from right to left and the boundary PEs having unique shapes and directions to complete the message pathway. Another important aspect of this diagram is that messages transiting between two adjacent routers are passed with 1-cycle latency, with each message carrying a 32-bit data payload. While adjacent routers communicate at 1-cycle latency, the latency between each processor and its router within the PE is much higher, taking 7 cycles.

\begin{figure}[h!]
\centering
\includegraphics[width=0.7\columnwidth]{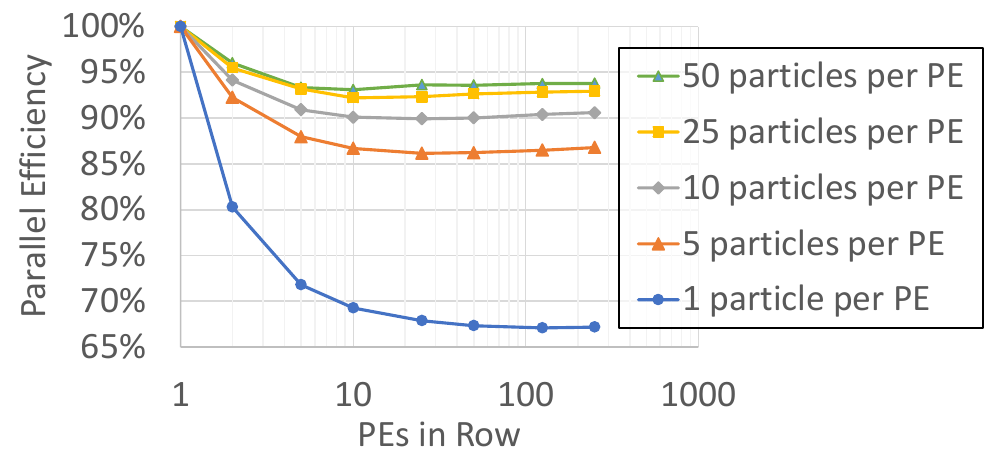}
\caption{Weak scaling for nuclide and particle decomposition across a row of PEs. The problem size per PE is fixed to hold one nuclide with 1,000 energy gridpoints and 5 cross section reaction channels and starts $n$ particles, with a variety of values for $n$ compared.}
\label{fig:row_weak} 
\end{figure}

We begin our analysis of the round-robin algorithm with a weak scaling study, wherein the problem size per PE is fixed. Our baseline problem size is for 1 nuclide per PE, with a nuclide having 1,000 energy gridpoints and 5 cross section reaction channels. Since it is desirable that each PE have at least one task to do (e.g., processing of at least a single particle for a single nuclide), we also pin the problem size per PE in a second dimension---the number of particles per PE, $n$. The sensitivity in performance to $n$ is also investigated. Weak scaling efficiency is then calculated separately for each study of $n$ by comparing the number of cycles per PE per nuclide per particle against the case where only 1 PE is used for that value of $n$. Our results, shown in \autoref{fig:row_weak}, show impressive weak scaling efficiency. With higher numbers of starting particles per PE, weak scaling out to a row width of 250 PEs can remain as high as 93\%. For the smallest problem sizes possible (where each PE has only a single nuclide and particle to operate on per communication round), the kernel was still reasonably efficient, at 67\% efficiency.

\begin{figure}[h!]
    \centering
    \includegraphics[width=0.7\columnwidth]{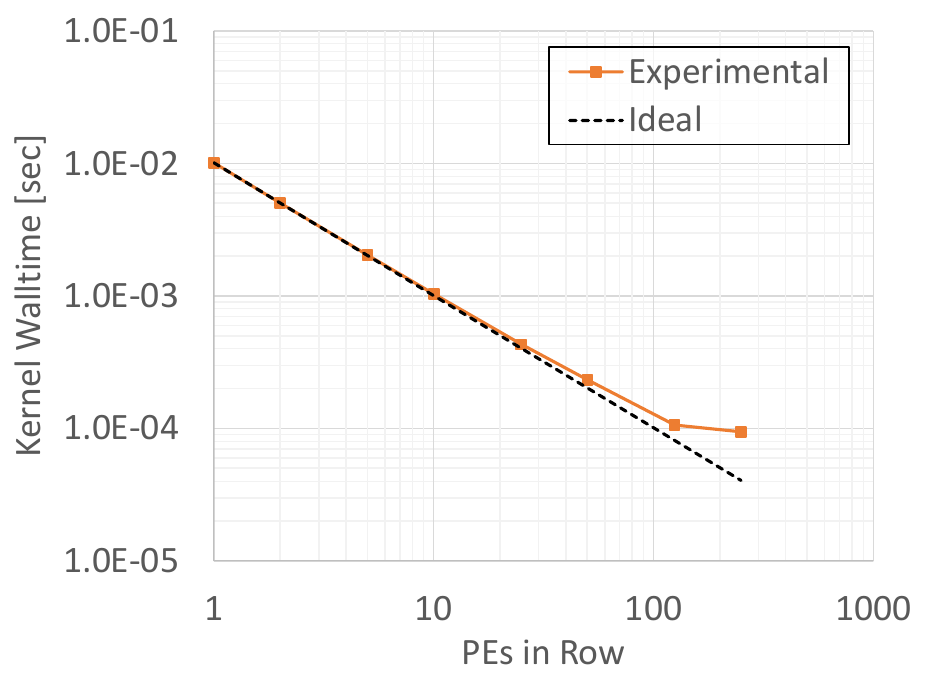}
    \caption{Strong scaling nuclide decomposition study across a row of PEs. The global problem size is fixed with 250 particles and 250 nuclides, with 10 energy grid points and one cross section reaction channel per nuclide.}
    \label{fig:row_strong}
\end{figure}

We also performed a strong scaling row study. In strong scaling, the global problem size is fixed; and as we add PEs to the row, we spread the static load out among more processors. This type of scaling is fairly unnatural on the WSE-2, given that it is difficult to fit a meaningful problem for 250 PEs using only the memory resources of a single PE. Thus, we use fairly minimal parameters for our global problem size, fixing the number of particles at 250, the number of nuclides at 250, the energy gridpoints per nuclide at 10, and with only a single cross section reaction channel. While the weak scaling analysis is likely to be more relevant in practice, a strong scaling analysis is nonetheless interesting given the added costs of communication. The results of our strong scaling analysis, given in \autoref{fig:row_strong}, show that the kernel performs surprisingly well in the strong scaling regime, with communication costs remaining trivial until about 125 PEs of width, after which point the communication costs begin to dominate. Given the results of the weak scaling study that indicated significant reductions in relative communication costs as the particle count was increased, we expect that a strong scaling analysis using more starting particles (which is not possible to fit into a single PE's memory for this problem) would theoretically show improved strong scaling performance.

\subsection{Load Balancing}
\label{sec:load_balance}

A major implication of our decomposition scheme is the potential for large stochastic load imbalances between the PEs of the WSE-2 grid. In particular, because of the random distribution of particle energy levels, the number of particles in each energy band (and within each PE of each energy band) will be uneven. While one can reasonably assume that the distribution will be approximately statistically uniform (assuming appropriate energy bands boundaries are selected), stochastic noise in this distribution may greatly inhibit the overall performance of the kernel. This is because overall kernel wall time will be dictated by the PE that starts with the largest number of particles after energy sorting. While these particles will be passed between each PE in the row as part of the round-robin exchange, in our current communication scheme this ``workgroup'' of particles will travel together and so will form a bottleneck regardless of which PE they travel to. Critically, this bottleneck is defined by the worst-case load peak out of any PE out of the approximately 750k usable PEs on the WSE-2. In other words, even for a particle distribution with fairly low noise, there are enough PEs that a high particle count outlier is inevitable. Therefore, we found that it was necessary to develop some mechanism for mitigating these load imbalance issues.

The goal for our load balancing technique is to ensure that peak load experienced by any PE is as close as possible to the average load. There are two dimensions upon which we can generally perform load balancing. The first is the number of particles that will be sampled to be within a given energy band (row). This determines the load imbalance between entire rows.
In theory, we could adjust the bounds of the energy band that a row holds on the fly after particle sorting, for instance by dynamically transferring nuclide gridpoint data and associated particles between rows based on the number of particles present in that row. This would be a complex communication pattern to implement, however, particularly given that it would involve some sort of full-row communication and synchronization to determine how to restructure the row's energy band.

The second dimension we can load balance is the number of starting particles held by the various PEs within a given row (energy band) after sorting. This is a much easier dimension to load balance over, because the particles within a row do not need to maintain any special ordering, so we are free to move the particles around as is desirable for load balancing. This would not be difficult to manage if we had a single process on each row that contained a full view of the sorted particle distribution across all PEs of the row, but the expense of accumulating this sort of data across all PEs (and then transmitting a remapping pattern back out to all PEs) may be prohibitively expensive. Rather, a more ideal pattern would not involve any global synchronization between PEs in a row, instead relying only on neighbor-to-neighbor communication. A final (and more practical) characteristic of a good load balancing pattern is that it should be simple to implement, given the complexity of implementing message-passing routines in the CSL language.

With these ideas in mind, we decided to forgo any between-row load balancing given the high degree of difficultly of reshuffling both particles and cross section energy band data between rows. Instead, we propose a ``diffusion-based'' load balancing technique to improve the distribution with each row independently. While focusing only on load imbalances within a row will mean that attainment of fully ideal load balancing will not be possible, it can at least transfer the bottleneck from the PE with the highest load (which is extremely sensitive to outliers in the starting distribution) to the row with the highest aggregate load (which, given that there are typically going to be tens or hundreds of PEs within each row, is a value that is likely to be much less noisy).

Our proposed ``diffusion-based'' load balancing technique applies an iterative process to the particle distributions within each 1D row. Similar to the round-robin row exchange phase, the diffusion phase is broken up into a number of communication iterations, where at each iteration each PE will transfer half of its current particles to its neighbor(s) and receive particles from its neighbors as well. At the asymptotic limit of many communication iterations, this diffusion operator will ensure that each row has a maximally uniform distribution of particles between its PEs. A great benefit of this strategy is that even just a few diffusion iterations should have a significant impact in the overall load balance, since the diffusion process tends to flatten peaks  quickly. While this diffusion process would optimally be implemented in a bidirectional manner, with PEs in a row able to send particles to both their left and right, for the sake of simplicity, we implement it in only a single direction so as to make full reuse of the round-robin communication pattern we have already implemented to handle the nuclide accumulation phase of the simulation. Thus, with only a few additional lines of code, we were able to add additional functionality into the round-robin row exchange routine so as to handle the particle diffusion phase as well. This makes the diffusion process a little less efficient (since PEs can  diffuse particles only to the right, with periodic boundary conditions), but nonetheless we expect that this will still greatly improve load balancing.

\begin{figure}[h!]
    \centering
    \includegraphics[width=0.7\columnwidth]{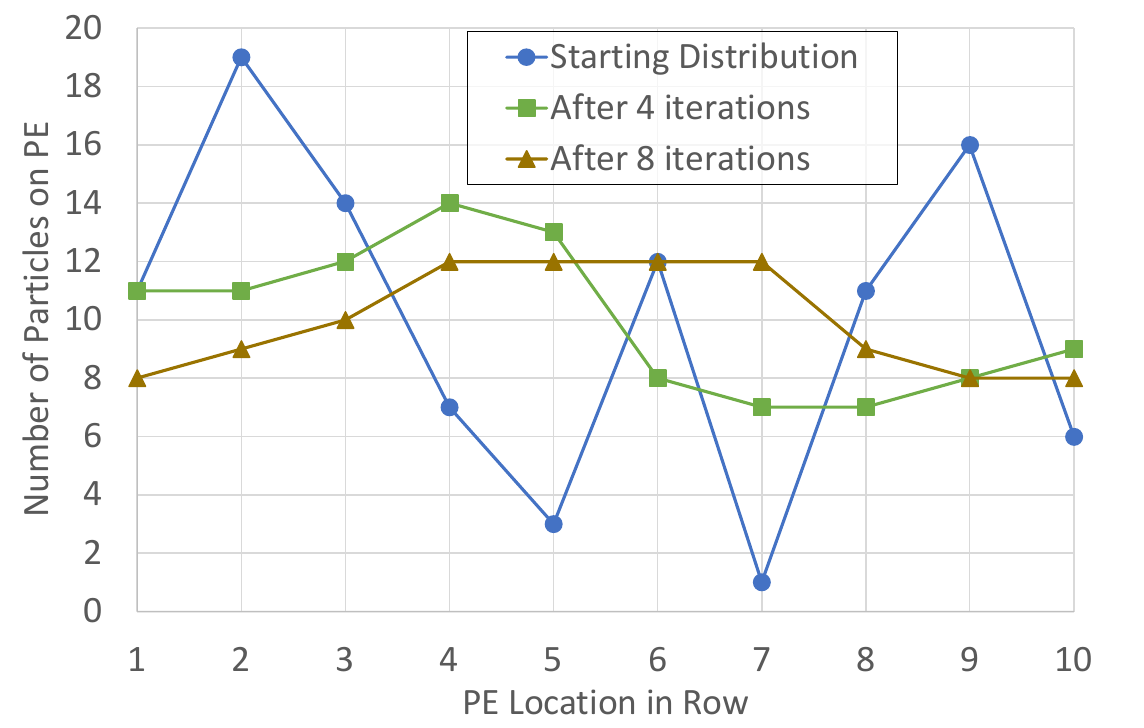}
    \caption{Simplified example of diffusion-based iterative load balancing process for a single row of 10 PEs.}
    \label{fig:diffusion_example}
\end{figure}

To demonstrate this process abstractly, we present an example that takes the case of a single row of PEs with 10 columns, with a randomized starting particle distribution. Each PE begins with a random number of particles between 0 and 20, and the single-direction diffusion process is then simulated, with the results of this process shown in \autoref{fig:diffusion_example}. In this example, the distribution begins with a PE holding 19 particles, creating a peaking load factor of 1.9x (meaning that a subsequent cross section lookup round-robin kernel would take about 1.9x longer than ideal to complete). After just 4 diffusion iterations, the particle distribution peak load has reduced to 14 (about 1.4x), and after 8 iterations, it has improved to a peak load of 12 particles (about 1.2x). 

\subsection{Tiling}

One natural limitation to our overall communication strategy is that our nuclide decomposition across PEs is naively limited by the number of physical nuclides being simulated in the problem. This would preclude utilization of the full resources of a WSE-2 (which is composed of a grid of $994 \times 750$ PEs), given that the highest number of nuclides present in a single material of a reactor simulation will be in the range of 250--300, well below the physical width of the WSE-2. Additionally, as we found in our column-scaling studies, the communication costs associated with sorting particles along a column length of 994 may be prohibitively costly.
Given these considerations, it is more efficient to decompose cross section data and particles into a PE subgrid of dimensions far smaller than the overall WSE-2 grid. This smaller subgrid can then be ``tiled'' and replicated to run concurrently (and independently) so as to fill up the entire WSE-2. Given that particle histories are independent, there is no need for communication across tile boundaries. We also note that only 60 MB of aggregate data is needed to represent our target problem nuclide cross section dataset (while the WSE-2 has ~40 GB of total memory), such that its replication across multiple tiles will not quickly exhaust the memory resources of the system. This tiling capability was added to our CSL implementation of the kernel, allowing the user to adjust the PE and tile configurations as inputs for the program.

\section{Full Cerebras WSE-2 Performance}
\label{sec:full_comms}

With communication schemes defined for sorting particles in energy, load balancing the particle load, moving particles between PEs for nuclide accumulation, and  replicating these processes within independent tiles so as to saturate the entire WSE-2, we now have all the elements needed to execute a fully decomposed cross -ection lookup kernel at scale. For filling the full WSE-2, we consider a variety of tile sizes that can accommodate the realistic target benchmark nuclide data size listed in \autoref{tab:xs_params} so as to determine the optimal configuration. In all cases, however, the simulated target problem is configured  to consistently represent a problem with 250 nuclides, about 10,000 energy gridpoints per nuclide, 5 cross section reaction channels, and 30 starting particles per PE, regardless of decomposition dimensions. The number of gridpoints per row and the number of nuclides per column were adjusted to conserve the global problem size for each configuration. 

For each tile configuration, we also ran three separate studies representing different assumptions regarding load balancing: ideal starting load distribution, realistic (random) starting load distribution, and realistic load distribution with our dynamic diffusion-based load balancing technique enabled.
In the ideal case, particles begin in a (seemingly) unsorted and random manner in each column, but we bias the sampling process to ensure that, after sorting, particles will end up in a perfectly load balanced distribution where every PE on the WSE-2 has the exact same number of particles. This biasing is accomplished by sampling particles for each row within that row's energy band and then shuffling the columns of particles independently on the host before transferring the particle data and launching the kernel on the device. Thus, this method still accurately captures the expense of sorting but excludes the possibility of load imbalance during the round-robin phase. The second case simply removes the biased sampling so that, after sorting, some PEs will have more particles than others. The third case is identical to the second except that the diffusion-based load balancing scheme (as discussed in \autoref{sec:load_balance}) is enabled.

The results of our full-machine analysis are shown in \autoref{tab:full_machine}. 
The first key finding is that the kernel tends to perform well on all of the three tested tile configurations, with the maximal and minimal ideal performance levels  varying only by about 10\%.
The second major finding is that, without use of a dynamic load balancing strategy, the random distribution of particles causes significant performance degradation. In all three tile configurations, each grid had at least one PE holding 1.8x more particles than average after the sorting operation was complete, as indicated by the ``Peak Load'' field of \autoref{tab:full_machine} that reports the maximum number of particles starting on any PE to the average number of particles started per PE. This load imbalance is problematic because the overall runtime of the round-robin nuclide accumulation phase will be dictated by whichever workgroup (i.e., a group of particles starting on one PE) is the largest, since this group will bottleneck the entire row as it migrates through the round-robin row exchange. The performance loss between the ideal and realistic starting condition is approximately the same as the load imbalance factor.

The third key finding was the success of our dynamic diffusion-based load balancing scheme. For all three configurations we used 100 diffusion iterations. Our diffusion-based load balancing stage added minimal overhead and greatly reduced the load peaking factors from 1.8x all the way down to 1.1--1.2x (very close to the ideal of 1.0). For the optimal $90 \times 125$ subgrid configuration, this resulted in a whole machine performance improvement of about 52\% and makes the kernel much less sensitive to noise in the starting particle distribution. While additional  gains may be possible by adding bidirectional diffusion or allowing for on the fly load balancing between rows, these changes would be much more complex to implement and would  result in only marginal gains because the current performance is only about 10\% slower than the ideal case with perfect load balancing.

\begin{table}[]
\centering
\caption{Full-machine results on a Cerebras WSE-2.}
\label{tab:full_machine}
\resizebox{\textwidth}{!}{%
\begin{tabular}{lllllllll}
\hline
\multicolumn{3}{c}{\begin{tabular}[c]{@{}c@{}}WSE-2 Grid\\ Configuration\end{tabular}} & \multicolumn{2}{c}{\begin{tabular}[c]{@{}c@{}}Ideal (Uniform)\\ Particle Distribution\end{tabular}} & \multicolumn{2}{c}{\begin{tabular}[c]{@{}c@{}}Realistic (Random)\\ Particle Distribution\end{tabular}} & \multicolumn{2}{c}{\begin{tabular}[c]{@{}c@{}}Realistic (Random)\\ Particle Distribution\\ with Diffusion Stage\end{tabular}} \\ \hline
\begin{tabular}[c]{@{}l@{}}Tile\\ Dimensions\end{tabular} & \begin{tabular}[c]{@{}l@{}}Grid\\ Dimensions\end{tabular} & \begin{tabular}[c]{@{}l@{}}Total\\ PEs\end{tabular} & \begin{tabular}[c]{@{}l@{}}Performance\\ {[}Lookups/s{]}\end{tabular} & \begin{tabular}[c]{@{}l@{}}Peak\\ Load\end{tabular} & \begin{tabular}[c]{@{}l@{}}Performance\\ {[}Lookups/s{]}\end{tabular} & \begin{tabular}[c]{@{}l@{}}Peak\\ Load\end{tabular} & \begin{tabular}[c]{@{}l@{}}Performance\\ {[}Lookups/s{]}\end{tabular} & \begin{tabular}[c]{@{}l@{}}Peak\\ Load\end{tabular} \\ \hline
124 PE x 25 PE & 8 Tiles x 30 Tiles & 744,000 & 9.13E+09 & 1 & 5.51E+09 & 1.8 & 7.76E+09 & 1.2 \\
90 PE x 125 PE & 11 Tiles x 6 Tiles & 742,500 & 9.25E+09 & 1 & 5.51E+09 & 1.8 & 8.36E+09 & 1.1 \\
62 PE x 250 PE & 16 Tiles x 3 Tiles & 744,000 & 8.42E+09 & 1 & 4.93E+09 & 1.8 & 7.63E+09 & 1.1 \\ \hline
\end{tabular}%
}
\end{table}

While our results in \autoref{tab:full_machine} show that the load imbalance between PEs can be successfully mitigated, it is unclear from this data alone what the overall communication costs were for the kernel as a whole. Thus, an important question to answer is what the overall communication overhead is for our decomposition, load balancing, and particle movement scheme combined. Are significant further gains in performance possible by developing more optimal communication patterns, or is the kernel performing close to the ideal? Given the fine-grained decomposition of data that is required to fit each subdomain within 48 kB of local PE memory, were communication costs prohibitively expensive? 

We can answer these questions by computing an idealized optimal cycle count based on extrapolating the performance values gathered when developing the single-PE optimized kernel. In \autoref{tab:pe_optimize} we found that our optimal single PE performance (with no communication routines) required 200 cycles per nuclide per particle. Thus, to process 30 particles with 250 nuclides (assuming 161 local energy gridpoints), we can compute the total cycle count as 250 cycles per nuclide per particle $\times$ 250 nuclides $\times$ 30 particles $=$ 1,875,000 cycles.

Our optimized full-machine WSE-2 run for the $90 \times 125$ PE case in \autoref{tab:full_machine} consumed 2,264,078 cycles (the maximum cycle count out of all PEs on the WSE-2). We can compare this value with the optimal single-PE (per-nuclide, per-particle) value of 1,875,000 cycles.

Thus, the full cost of our communication scheme  adds only 21\% overhead as compared with an ideal case (assuming each PE had infinite memory and hence  could replicate the full cross section dataset locally) where no communication was required. Additionally, we note that about half of this cost can likely be attributed to the remaining 10\% load imbalance. In this light, the opportunity for further optimizations to our decomposition and communication strategies (as well as their implementation details in CSL) appears to be  narrow given the excellent performance that has already been achieved with this scheme.

\section{Comparison with  GPU}

In order to appraise the performance of our XS lookup kernel on the Cerebras WSE-2 machine, a baseline was needed. Recent work in  Monte Carlo particle transport has shown that GPUs tend to be far more efficient for this algorithm than are CPUs~\cite{Tramm2022,hamilton2019ane,pragma}. We therefore decided that the A100 GPU would make  an ideal representative for what is possible for performance of this algorithm on more mainstream HPC architectures. In particular, the A100 is a good candidate for comparison with the WSE-2 given that both chips were manufactured by Taiwan Semiconductor Manufacturing Company using a 7 nm process. To this end, we have ported the kernel into CUDA and have considered a variety of GPU- and CUDA-specific optimizations to ensure that a fair comparison is made. In other words, we take a ``gloves off'' approach and consider all possible optimizations for the GPU (just as we did with the WSE-2), even if the optimizations do not translate between architectures.

Before any performance comparisons are made, it is important to set reasonable expectations given the significant resource disparities between the wafer-scale WSE-2 and the more traditionally sized A100 chip. This is an important comparison to make because, given the trouble of developing code in Cerebras's proprietary CSL programming model and the difficulty of developing performant decomposition and communication schemes, we would hope that the WSE-2 would be able to offer some performance advantage over an equivalently scaled cluster of A100 GPUs. A few of the possible ways of making a fair comparison are given in \autoref{tab:resource_compare}.\footnote{While comparison of transistor counts, die area, and power are straightforward, comparison of peak floating-point capability is more nuanced. A single WSE-2 PE can do up to one FP32 fused multiply-add per cycle or two FP32 floating-point adds per cycle, in either case resulting in two FP32 operations per cycle. The clock rate of the Argonne CS-2 installation used in this study is 850 MHz, although the WSE-2 is capable of clocking up to 1 GHz (though this may cause an increase in thermal throttling). If excluding system and memory operation reserved PEs, such that only 745,500 PEs are used out of the 850,000 total, then we compute the theoretical maximum performance of the WSE-2 as $750 \times 994 \times 2 \times 850 \times 10^{6} = 1.267 \times 10^{15}$ FP32 FLOPS. If using a more liberal interpretation of theoretical performance (i.e., using all PEs and assuming a steady 1 GHz clock), the value would increase to 1.7 PFLOPS.}

\renewcommand{\thefootnote}{\alph{footnote}}

\begin{table}[h]
\centering
\caption{Comparison of an NVIDIA A100 (SXM4 40GB) and Cerebras WSE-2 architectures.}
\label{tab:resource_compare}
\begin{tabular}{@{}llllll@{}}
\toprule
 & \begin{tabular}[c]{@{}l@{}}Transistor\\ Count\\ {[}Trillion{]}\end{tabular} & \begin{tabular}[c]{@{}l@{}}Die Area\\ {[}mm$^2${]}\end{tabular} & \begin{tabular}[c]{@{}l@{}}Peak\\ Power\\ {[}kW{]}\end{tabular} & \begin{tabular}[c]{@{}l@{}}Theoretical\\ FP32 Peak\\ {[}TFLOPS{]}\end{tabular} & \begin{tabular}[c]{@{}l@{}}Monte Carlo\\ XS Lookup FOM\\ {[}Lookups/s{]}\end{tabular} \\ \midrule
A100 GPU & 0.0542 & 826 & 0.4 & 19.5 & 6.43E+07 \\
Cerebras WSE-2 & 2.6 & 46,225 & 22.8 & 1,267 & 8.36E+09 \\
WSE-2/A100 & 48 & 56 & 57 & 65 & 130 \\ \bottomrule
\end{tabular}
\end{table}

Since the A100 GPU we used for our testing featured 40 GB of high-bandwidth memory, there is no explicit need for cross section data decomposition. This greatly simplifies the implementation, although a few fine-grained optimizations are still considered. In particular, we tested several of the same optimization strategies used to optimize the kernel for the WSE-2  (as described in \autoref{sec:single_pe_opt}), as well as other techniques that make sense only in the context of GPUs.

The first optimization for the GPU we considered was use of half-precision arithmetic for the division operation in the interpolation phase of the lookup kernel. The second optimization was to consider use of stochastic interpolation instead of linear interpolation. To facilitate this, we used two different methods for pseudorandom number generation---NVIDIA's first-party cuRAND library and a minimal linear congruential generator (LCG). We note that NVIDIA's cuRAND library is simply an optimized software library; there are no special hardware units on the A100 for random number generation. The third optimization was to sort particles before running the lookup kernel. This optimization was easy to implement via the NVIDIA CUDA Thrust sorting library. Timing results presented for this optimization include the costs of sorting, which were fairly small compared with the cost of the cross section lookup kernel itself.

An additional optimization was also implemented that is not practical on the WSE-2 given per-PE memory constraints. This optimization, known as the ``double indexing'' or as the ``unionized energy grid (UEG)'' algorithm~\cite{Leppanen2009}, is an acceleration technique that functions by reducing the number of binary searches required for each XS lookup. The key optimization that this scheme makes is that only one binary search is required, rather than one binary search for each nuclide's energy grid, as is typically needed. For example, for a material with 250 nuclides, 250 binary searches would normally be required, but with the UEG approach only a single binary search is needed. The downside to this optimization strategy is that a significant additional quantity of memory is required to build a large table of indices. In order to implement the UEG optimization, a ``unionized'' grid of all energy levels of all nuclides is generated and sorted. For real cross section data, some nuclides will have some overlap between energy points; however, this overlap is typically small ($<20$\%), so for our synthetic data we will assume that all nuclides feature unique energy grid points. This means that the total number of energy gridpoints in the unionized energy grid will be equal to the number of nuclides times the number of gridpoints per nuclide. For each gridpoint on the unionized energy grid, we store an array with an index into each nuclide's energy grid corresponding to the energy level that is at or just below the unionized energy gridpoint. At kernel runtime, a thread will perform a single search on the unionized energy grid, and it will then have a map of where the corresponding energy level can be found in each nuclide's grid. Theoretically, all cross section data can be stored on the unionized grid as well, although this results in a lot of replicated data, and given that 5 or more reaction channels are typically stored, this can result in an impractically large amount of data. Overall, for a typical problem of 250 nuclides and 10,000 gridpoints per nuclide, the addition of the UEG acceleration structures adds about 2.5 GB of memory usage. 

The UEG approach is theoretically possible to implement on the WSE-2, although it greatly increases per-PE memory requirements. When fully decomposed in energy such that each row of PEs on a WSE-2  holds only 10 energy gridpoints per nuclide, for 250 nuclides this would increase the per-PE storage requirements from  needing to store only 10 nuclide energy gridpoints with 5 reaction channels (about 240 bytes) up to needing to store $10 \times 250 = 2500$ energy gridpoints and an equivalent number of indices (about 15 kB in total). While this can be reasonably fit within the 48 KB of memory per PE, it would preclude the extensive use of tiling to reduce communication costs, since at most two to three tiles could be used in the energy dimension, which would significantly increase overall communication costs.

We also considered an ``energy-banding'' approach on the GPU where smaller bands of cross section data are imported sequentially into shared GPU memory for faster access. However, we found that particle sorting tended to make this sort of optimization unnecessary because, when sorted, all particles within a warp tended to access the same exact global memory data. In this case, movement of global memory into local memory would not be expected to be amortized with any reuse.

The results of our optimization studies on the GPU are shown in \autoref{tab:gpu_optimize}. The primary optimization on the GPU was clearly the sorting of particles, which allows threads within a warp to access the same cross section data. Once particles were sorted, other optimizations had  little impact. For instance, use of stochastic interpolation and the unionized energy grid  improved sorted performance by only 1\%. Overall, the relevant figure of merit for a single A100 GPU with a maximally optimized kernel implementation is 64.3 million lookups/sec. We compare this value with the measured full-machine value from the Cerebras WSE-2 in \autoref{tab:resource_compare} and find that the WSE-2 achieved a rate of 8.36 billion lookups/sec with realistic load balancing. Therefore, the WSE-2 was about \SPEEDUP{}x faster than a single A100 GPU.

\begin{table}[h!]
\centering
\caption{Cross section lookup kernel optimizations for an A100 (40 GB SMX4) GPU/}
\label{tab:gpu_optimize}
\begin{tabular}{@{}lll@{}}
\toprule
 & \begin{tabular}[c]{@{}l@{}}Lookups\\ per sec\end{tabular} & \begin{tabular}[c]{@{}l@{}}Speedup\\ over\\ Baseline\end{tabular} \\ \midrule
Baseline & 2.94E+07 & - \\
Sorting & 6.33E+07 & 2.16 \\
Sorting + FP16 Division & 6.31E+07 & 2.15 \\
Sorting + Stochastic Interpolation (LCG) & 6.37E+07 & 2.17 \\
Sorting + Stochastic Interpolation (cuRAND) & 6.22E+07 & 2.12 \\
Sorting + Unionized Energy Grid & 6.41E+07 & 2.18 \\
Sorting + Unionized Energy Grid + LCG & 6.43E+07 & 2.19 \\ \bottomrule
\end{tabular}%
\end{table}

Our analysis of GPU performance has so far allowed each architecture to be run in an architecture-specific configuration where the optimal problem size (e.g., number of particles) is used for each architecture. For instance, the full-machine WSE-2 performance results given in \autoref{tab:resource_compare} correspond to a total problem size of about 22.3 million particles. The A100 GPU utilized a total problem size of 100 million particles. This is notable because the Cerebras is about 50x larger than the GPU, yet was running with high efficiency on a problem size 4.5x smaller than what was needed to saturate a single A100 GPU. This is an  important point of divergence between the two architectures and an area where the WSE-2 tends to stand out. While GPUs require a massive amount of parallelism to be expressed in order to allow for full masking of latency to main memory, the WSE-2 architecture can achieve reasonable efficiency with even only a single starting particle per PE.

To more fully understand the relative differences in small problem size performance, we ask a simple question: For a problem size of 744,000 particles running on the WSE-2, how many GPUs would be needed to strong scale to in order to be able to match the WSE-2's wall time? In this case, not even an infinite number of GPUs would be able to match the WSE-2's performance when considering small problems like this. This is shown clearly by the asymptotic strong scaling limit of GPU performance, where we consider the wall time it takes for the GPU to process just a single particle in isolation. The kernel time on an A100 is about 368 $\mu$s to process the single particle (with sorting overhead excluded, since it would not be needed if just a single particle were used). Comparatively, the wall time it takes for a WSE-2 to process 744,000 particles (i.e., one starting particle per PE) is only 280 $\mu$s (237,914 cycles), including all communication costs. Thus, we highlight that while the WSE-2 is shown to be \SPEEDUP{}x faster than a single A100, it may often take far more than \SPEEDUP{} A100 GPUs to match the performance of a single WSE-2 on all but the largest problem sizes, given the A100's loss of efficiency when strong scaling.

\section{Comparison to Simulated ASIC}

Recently, a custom application-specific integrated circuit (ASIC) for the MC cross section lookup kernel was designed~\cite{xsbench-chisel} using the Chisel hardware construction language~\cite{Bachrach:ft}. The total latency of the kernel was analytically estimated. Its resource usage (e.g., gate counts) was estimated based on Verilog code generated from the Chisel description of the kernel. The study found that the per-nuclide latency of a moderately optimized ASIC implementation of the kernel was 46 cycles and required an area of about 0.046 $mm^2$ if using a 14\,nm node process. The ASIC implementation uses 32-bit floating-point arithmetic operations. If such an architecture were to be naively replicated so as to fill up a chip the size of a Xeon 8180M CPU (also a 14\,nm architecture, having about 8 billion transistors), the paper estimated that approximately 326 billion nuclides per second could be processed at 1\,GHz, corresponding to about 1.304 billion lookups per second if assuming 250 nuclides are required. While these values ignore some practical complexities that may reduce the per-transistor efficiency of a production ASIC (e.g., handling of I/O on and off the chip), the study was nonetheless very useful as it serves as a ``high water mark'' for what level of performance is theoretically possible for this kernel. As the number of transistors in the 8180M, A100 GPU, and WSE-2 is known, we can generate transistor-normalized estimates of performance so as to get a sense of how efficient these architectures are as compared to the ASIC kernel implementation. \autoref{tab:ideal}\footnote{While comparing an ASIC implementation to code running on instruction-set architecture is not a like-for-like comparison, we provide the table to assess the approximate efficiency of accelerators using an ASIC reference as a baseline.} reveals that while the WSE-2 is clearly much more efficient than the A100 architecture for this algorithm, there is still potential for improvement in hardware design to better accommodate the needs of the cross section lookup algorithm.  

\begin{table}[]
\centering
\caption{Comparison of cross section lookup kernel to an ASIC kernel implementation.}
\label{tab:ideal}
\resizebox{\textwidth}{!}{%
\begin{tabular}{@{}lllll@{}}
\toprule
 & \begin{tabular}[c]{@{}l@{}}Performance\\ {[}lookups/sec{]}\end{tabular} & Transistors & \begin{tabular}[c]{@{}l@{}}Normalized\\ Performance\\ {[}lookups/sec/transistor{]}\end{tabular} & \begin{tabular}[c]{@{}l@{}}Architecture\\ Efficiency\\ {[}\% of baseline{]}\end{tabular} \\ \midrule
ASIC baseline & 1.30E+09 & 8.00E+09 & 1.63E-01 & 100.00\% \\
WSE-2 & 8.36E+09 & 2.60E+12 & 3.22E-03 & 1.97\% \\
A100 GPU & 6.43E+07 & 5.42E+10 & 1.19E-03 & 0.73\% \\ \bottomrule
\end{tabular}%
}
\end{table}

\section{Future Work}

\subsection{Prospects for a Full Monte Carlo Application on the WSE-2}

Our analysis has covered only  a single kernel from the Monte Carlo particle transport algorithm. While it is typically the most expensive kernel in the MC algorithm (at least in the context of nuclear reactor simulation problems), significant additional research would be required in order to implement a fully featured Monte Carlo application on the WSE-2. For a full-physics MC simulation to be performed on a WSE-2, several additional kernels would need to be implemented to complete the particle transport loop, and several additions would need to be made to the cross section lookup kernel as well due to simplifications that were made in this analysis.

The major simplifications made in this analysis for the MC cross section lookup kernel were as follows:

\begin{itemize}
    \item Use of only single-temperature cross section data. A realistic simulation will likely need to handle multiple XS datasets, each at a different temperature, with stochastic interpolation performed between the levels.
    \item Lack of S$(\alpha,\beta)$ thermal scattering data. A realistic simulation will need to store a small amount of additional data for certain nuclides in this low-energy range.
    \item Lack of probability tables in the unresolved resonance range. A realistic simulation will need to store a small amount of additional data in this high-energy range.
    \item Representation of only a single material. A realistic simulation will need to handle multiple material types, each with their own isotopic compositions. 
\end{itemize}

We believe that these missing capabilities should be feasible to add  to the kernel without significantly changing the fundamental decomposition scheme or particle exchange routines. For instance, use of multiple temperature levels may  require only that slightly larger tiles be used in order to reduce the per-PE memory overhead of storing nuclide XS datasets for multiple temperature levels.

Implementation of the remaining kernels that form the basic Monte Carlo particle transport loop would undoubtedly require additional research and  the development of new algorithms. However, the two other fundamental kernels that are required (ray tracing and collision physics) do not require as as much baseload data compared to what is typically required for cross section storage, such that data replication may be feasible. While spatial domain decomposition would likely be required for highly complex and unstructured geometries featuring millions of unique cell definitions, many simulation problems could be easily fit within the memory limitations of a WSE-2 PE. For instance, full-core nuclear reactors often feature lattice-based geometries, where only a handful of basic pin cell universes are defined and then repeated in nested lattice structures that define the entire reactor geometry. In these instances,  few surfaces and constructive solid geometry cells actually need to be stored in memory, such that it is feasible to define the full reactor geometry within only a few kilobytes of data. In cases where the full geometry cannot be replicated across all PEs, it may still be feasible to domain decompose only across tiles, rather than across all PEs. In this paper's CSL implementation, we replicated everything between tiles, but in theory it would be possible to domain decompose the simulation geometry across tiles and to add a reshuffling stage to allow particles to be exchanged between tiles when necessary.

\subsection{Alternative Algorithms}

In addition to the algorithms described so far in this paper for decomposing cross section data across the WSE-2 grid and for moving particles through the network for processing, we also considered several alternative algorithms. Some of these algorithms we note here as topics for future work, while others were tested and were found to be less performant than the algorithms so far discussed in this paper.

\begin{figure}[h!]
    \centering
    \includegraphics[width=0.7\columnwidth]{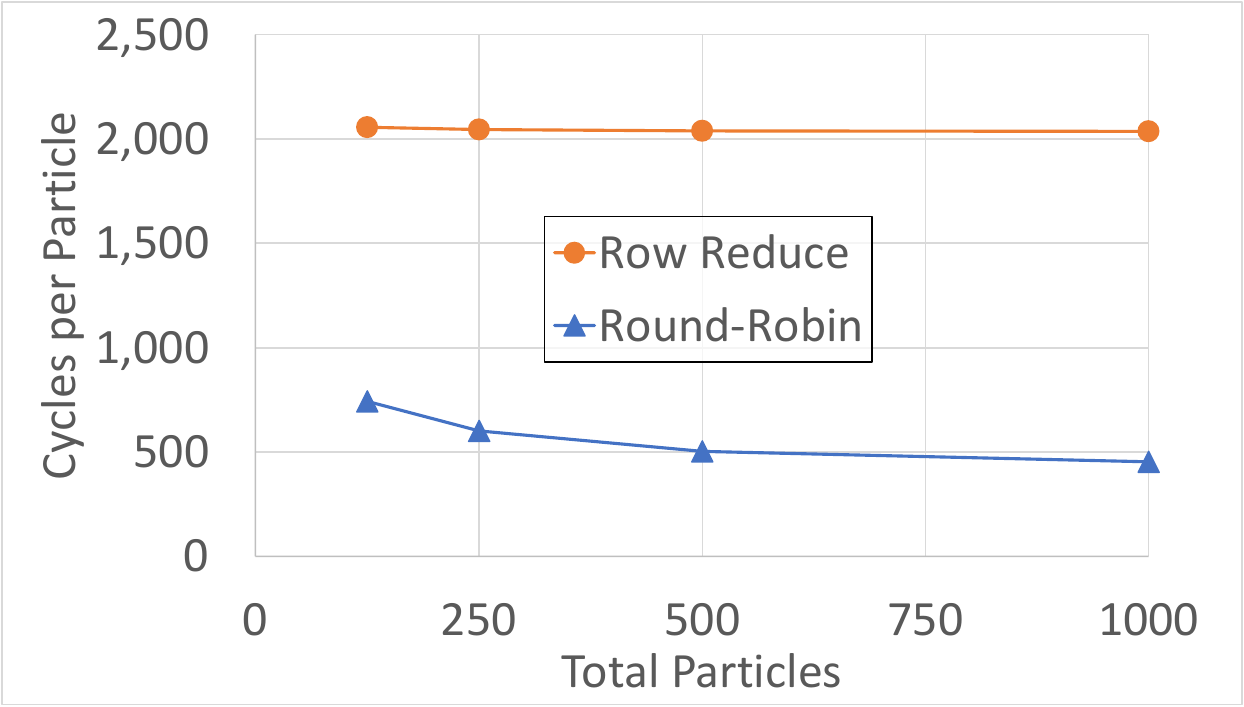}
    \caption{Comparison of the row-reduce algorithm with the round-robin algorithm. This study uses a single row of 125 PEs, with 125 nuclides, 10 gridpoints per nuclide, and 5 reaction channels in total, with the number of total particles varied. Cycle counts per particle per nuclide are reported, with both methods using traditional linear interpolation and without use of vector intrinsics.}
    \label{fig:row_reduce}
\end{figure}

We first consider an alternative to the method for performing the row exchanges for accumulating nuclide information defined in \autoref{sec:row_nuclide_accum}. In the alternative ``row-reduction'' method, nuclides are decomposed across the PEs in a row (just as in our original ``round-robin'' method), but particles begin by being copied to all PEs. Each PE then performs a lookup for its local nuclides for all the particles at once, and then a row reduction operation is performed. The immediate advantage to this method is that it leverages the CSL row reduction abstraction, meaning that the communication pattern between PEs within a row does not need to be manually programmed or coordinated. As shown in \autoref{fig:row_reduce}, however, this method resulted in an increase of 2.7--4.5x in overall runtime costs as compared with the round-robin approach, such that the added complexity of the round-robin did appear to be unavoidable. The significant difference in cost was likely due to the row-reduction method not being able to mask communication costs with useful work, since all work had to be completed up-front before communication could begin. Another downside to the row-reduction method is that particles must be copied to all PEs in the row before the algorithm can begin; and ultimately particles end up reduced on only a single PE of the row, which does not map naturally to the event-based algorithm that this kernel would likely be used with if extended to a full-physics implementation. Conversely, the round-robin approach does not require that particles be copied, and at its conclusion particles are still evenly distributed throughout the row, which allows for other kernels (e.g., ray tracing, collision physics, tallying) to be called in an event-based algorithm without having to necessarily reorganize the particles again.

Another key idea we developed for further cutting communication costs when many particles are used (or when particle objects are very large) is to flow the cross section data through the round-robin row exchange instead of the particle data. This would seem to be advantageous when the total nuclide cross section data per PE is less than the total particle buffer size, as might be the case when simulating many particles at once. However, it is left for future work to experiment with this algorithm.

Additionally, it may be advantageous to decompose only in energy space and to simultaneously decompose energy across the global 2D grid. That is, instead of decomposing by energy in one dimension and by nuclide in another dimension, each PE would hold data for all nuclides but would hold only a very narrow slice of energy (potentially, just three or four energy points per nuclide). Particles would then be tasked with sorting themselves into the correct band in two dimensions before computation begins. This algorithm is also left for future work for testing.

It also may be possible to load balance across rows dynamically in response to particle energy distributions. For instance, a row having twice as many particles as its neighbor might shift a portion of its cross section data to its neighbor along with particles within that energy space in order to even the load.

A final optimization that we consider for future work is to apply the fractional cascading~\cite{fractional} technique for accelerating binary search operations on sequential nuclide lookup operations. Our kernel currently performs a full (naive) binary search on each nuclide. The fractional cascading technique performs a preprocessing step before the simulation begins where, for each energy grid point in nuclide $n$, the closest gridpoints above and below this energy level in the grid of nuclide $n+1$ are determined and stored locally. Thus, after the first nuclide has been processed, the search space for subsequent nuclides is often greatly limited, reducing the number of binary search steps needed. The downside is that particles must carry this indexing information with them as they travel between PEs in a row, and that more memory is used to store the additional indexing data for each nuclide.

\section{Conclusions}

In this study we ported a simplified version of the Monte Carlo cross section lookup kernel using the Cerebras SDK and the Cerebras CSL programming model and evaluated the performance of the kernel on the Cerebras WSE-2 wafer-scale machine. Beyond the challenge of porting the kernel into the low-level CSL programming model, a number of new algorithms were proposed and tested to handle the decomposition of cross section data into the small 48 kB local memory domains that each of the WSE-2's approximately 750,000 PEs contains. We also had to develop several algorithms to sort particles in energy space, load balance them, and then flow them through portions of the WSE-2 so as to accumulate all required cross section data.

Our decomposition and communication scheme involved three  stages: (1) the sorting of particles into energy bands within each column of PEs, (2) an iterative diffusion-based load balancing stage for balancing starting particle loads within each row, and (3) a row-wise round-robin exchange of particles to allow particles to accumulate nuclide information from each column in the row. Importantly, all of these communication patterns had to be developed to avoid any concept of global synchronization or point-to-point message passing, given the limitations of the WSE-2 hardware. Each of the communication patterns is limited to only neighbor-to-neighbor exchanges within the 2D grid of PEs that composes the WSE-2.

In addition to these communication patterns, we  developed an architecture-specific optimization that leverages the unique hardware capabilities of the WSE-2. In particular, each PE of the WSE-2 has specialized silicon dedicated to the generation of random numbers (in support of stochastic gradient descent and other common stochastic machine learning algorithms). We found we were able to leverage this feature to replace an expensive linear interpolation operation (which involves a very expensive floating-point division) with a stochastic interpolation scheme that improved overall lookup kernel performance by over 65\%.

When our algorithm was run at scale on a full Cerebras WSE-2 chip, we found that our dynamic load balancing scheme was able to successfully flatten peak per-PE load factors from 1.8x down to 1.1--1.2x, resulting in a similar factor of speedup. We also found that the total communication cost of our decomposition scheme was only about 21\%,  which is surprisingly low given the fine-grained nature of our data decomposition across the nearly 750,000 PEs of the WSE-2.

To provide a baseline to contextualize the performance of the WSE-2, we also developed a highly optimized CUDA kernel for testing on an A100 GPU. We implemented stochastic interpolation optimizations on the GPU; but because of the A100's lack of dedicated random number generation hardware, these strategies were not found to be helpful there. However, other GPU-specific optimizations (like particle sorting via CUDA thrust, implementing a memory-expensive unionized energy grid, and investigating the use of shared memory) were implemented to ensure that the GPU kernel was maximally optimized given the specific capabilities of the NVIDIA architecture.

Overall, we found that a single Cerebras WSE-2 wafer-scale chip was about \SPEEDUP{} times faster than a single A100 GPU, when both systems were run using system-optimal problem sizes (particle counts). This result is significantly more than expected given the 48--65x relative difference of silicon provisions and power metrics of the systems. 

Was this exercise worthwhile then? Clearly, the WSE-2 provides a significant speedup over the A100, even exceeding expectations given the provisions of the two systems. Arguably, the increase in performance did come at a cost---namely, increases in both software programming and algorithmic complexity. However, we also note that similar statements could be said about GPU general-purpose programming when it was in its infancy. In light of how AI accelerators such as the WSE-2 were designed almost exclusively around deep learning AI tasks, it is noteworthy that the WSE-2 is already able to exceed performance expectations relative to GPUs---an architecture that has had several decades to mature and that is now quite friendly to HPC simulation applications. One can imagine that relatively small hardware design changes might be made to future Cerebras architectures (and other AI-centric accelerators) that may further improve the performance of simulation applications on them and that new programming models might be developed to target these architectures in a higher-level and more portable manner. In this light, we believe that algorithmic design and optimization for these architectures will be an important topic in the field of HPC simulation going forward.

All code written to support this paper is open source and is available at~\cite{mc_csl_zenodo}.

\section*{Acknowledgements}

The submitted manuscript was supported by the U.S. Department of Energy, Office of Science, under contract No. DE-AC02-06CH11357. This research used resources of the Argonne Leadership Computing Facility, a U.S. Department of Energy (DOE) Office of Science user facility at Argonne National Laboratory, and is based on research supported by the U.S. DOE Office of Science-Advanced Scientific Computing Research Program, under Contract No. DE-AC02-06CH11357.

\bibliographystyle{elsarticle-num} 
\bibliography{references}

\noindent\fbox{%
  \parbox{\textwidth}{\small %
    The submitted manuscript has been created by UChicago Argonne, LLC, Operator
    of Argonne \mbox{National} Laboratory (``Argonne'').  Argonne, a
    U.S. Department of Energy Office of Science laboratory, is operated
    under Contract No. \mbox{DE-AC02-06CH11357}.  The U.S. Government retains for
    itself, and others acting on its behalf, a paid-up nonexclusive, irrevocable
    worldwide license in said article to reproduce, prepare derivative works,
    distribute copies to the public, and perform publicly and display publicly,
    by or on behalf of the Government. The Department of Energy will provide
    public access to these results of federally sponsored research in accordance
    with the DOE Public Access
    Plan. \url{http://energy.gov/downloads/doe-public-access-plan.}
  }%
}

\end{document}